\UseRawInputEncoding
\documentclass[preprint,
 amsmath,amssymb,
 aps,
]{revtex4-2}

\usepackage{graphicx}% Include figure files
\usepackage{dcolumn}% Align table columns on decimal point
\usepackage{bm}% bold math
\usepackage{subcaption}
\usepackage{xcolor}
\usepackage{caption}
\usepackage{hyperref}
\usepackage[utf8]{inputenc} % Allows native accent input (Default in modern Overleaf)
\usepackage[T1]{fontenc}    % Ensures accents copy/paste cleanly from the output PDF
\usepackage[english]{babel}% Handles local hyphenation (change 'spanish' to your language)
\begin{document}

\preprint{This manuscript has been submitted to Physical Review E for review}
\title{Iterative detection of global factors near the BBP phase transition}

\author{Andr\'es Garc\'ia-Medina}
\email{andgarm.n@gmail.com}
 \affiliation{Tijuana Institute of Technology, Tecnológico Nacional de M\'exico, Calz. del Tecnol\'ogico 12950, Tom\'as Aquino, Tijuana, 22414, Baja California, M\'exico}%

\date{\today}

\begin{abstract}
Detecting the number of global factors in high-dimensional correlation matrices is a central problem in multivariate statistics and random matrix theory, with important implications for asset pricing and econophysics. When the number of variables $p$ is comparable to the number of observations $n$, signal-to-noise separation becomes difficult, especially near the Baik--Ben Arous--Péché (BBP) transition, where weak factors may be confused with fluctuations at the Marčenko--Pastur spectral edge. In this work, we characterize the participation-ratio (PR) structure of the Brown--Harding (BH) factor model. Under strong common loadings, the leading coherent eigenvector $u_1$ satisfies $\mathrm{PR}(u_1)/p\to 1$, whereas weak-factor directions and typical idiosyncratic sample eigenvectors $u$ satisfy the delocalized benchmark $\mathrm{PR}(u)/p\to 1/3$. These limits motivate an eigenvector-level criterion for retaining extensive directions. Building on this result, we propose an iterative global factor (IGF) algorithm that combines adaptive Marčenko--Pastur edge recalibration with a PR delocalization filter. The method iteratively reestimates the effective noise level, tests eigenvalue separation from the residual bulk, and retains only spectrally separated components with sufficiently extended eigenvectors. Monte Carlo simulations of the BH factor model show that IGF recovers the true number of factors near the BBP transition, where eigenvalue-only criteria, including the Tracy--Widom and Onatski tests, can fail or remain ambiguous. A synthetic moving-window calibration matched to the empirical dimensions provides the operational threshold $\tau=0.3$, which is then applied to S\&P 500 returns. In the empirical analysis, IGF detects a richer and more dynamic set of global factors than the Onatski test, with a median count of 7 factors across moving windows. The results indicate that combining spectral separation with eigenvector delocalization improves the detectability, stability, and interpretability of global-factor estimation in high-dimensional financial correlation matrices.
\end{abstract}
\maketitle

\noindent\textbf{Keywords:} random matrix theory; high-dimensional correlation matrices; BBP phase transition; econophysics; asset-pricing models. 

\noindent\textbf{PACS:} 02.10.Yn; 02.50.-r; 89.65.Gh; 89.75.-k.

\section{Introduction}
Determining the number of relevant factors is crucial in finance for asset valuation, investment decisions, and the understanding of asset-pricing mechanisms. Several heuristics and estimators have been proposed for this purpose in classical statistical settings \cite{cattell1966scree,wold1978cross,zientek2007applying,kaiser1960application,guttman1954some}. Nevertheless, the high-dimensional setting, in which the number of variables, such as assets or markets, is of the same order as the number of observations, such as trading days, remains challenging and has become an active area of research.

This regime is also of particular interest in econophysics for two main reasons. First, the phenomenological approach of physics to finance focuses on stylized facts observed in financial markets, including the rich dynamics that emerge from the complex interactions of many assets over short time intervals. Second, random matrix theory (RMT), originally developed in physics to describe complex many-body systems, provides a natural mathematical framework for studying high-dimensional financial correlation matrices \cite{mehta2004random}. From a financial and statistical perspective, the high-dimensional regime is also relevant for portfolio diversification and for reducing the effects of nonstationarity in time series.

A recent RMT-based proposal for estimating the number of factors in high-dimensional financial covariance matrices was introduced in Ref.~\cite{molero2023market}. The core of that method is based on the Onatski test \cite{onatski2008tracy,onatski2009testing}. Onatski \cite{onatski2010determining} showed that this criterion improves upon the Bai--Ng approach \cite{bai2002determining} in the presence of nontrivial correlations, reducing the tendency to overestimate the number of factors. However, weak factors may not be detected by such estimators because they can be easily confused with idiosyncratic components. Onatski \cite{onatski2012asymptotics} further extended the approximate-factor framework to a weakly influential factor regime, allowing for idiosyncratic dependence structures related to those considered by Bai and Ng \cite{bai2002determining}. In that regime, the total number of factors is not generally identifiable; instead, one can identify an effective number of factors whose principal-component estimates do not become asymptotically orthogonal to the true factors, using the edge-distribution procedure of Ref.~\cite{onatski2010determining}.

An additional difficulty arises from the Baik--Ben Arous--Péché (BBP) phase transition\cite{baik2005phase}, originally established for the largest eigenvalue of a complex spiked sample covariance matrix. The authors showed that a finite-rank population perturbation becomes spectrally detectable only when the corresponding population eigenvalue exceeds a critical threshold in the spiked covariance setting introduced by Johnstone \cite{johnstone2001distribution}. The extension to real sample covariance matrices was obtained by Baik and Silverstein \cite{baik2006eigenvalues}, while Paul \cite{paul2007asymptotics} studied the corresponding real Gaussian eigenstructure. Benaych-Georges and Nadakuditi \cite{benaych2011eigenvalues,benaych2012singular} extended the BBP-type phase-transition framework from classical spiked random matrix models to more general finite-rank perturbations of large random matrices, including additive, multiplicative, and rectangular signal-plus-noise settings.

In the context of factor models, Harding \cite{harding2008explaining} highlighted the bias that can arise when factor strength is inferred from leading sample eigenvalues in high-dimensional spiked systems and related the detectability of weak factors to the BBP phase transition through finite-sample corrections of the Brown factor model\cite{brown1989number}. Harding \cite{harding2007essays} and Yeo and Papanicolaou \cite{yeo2016random} also explored the use of free-probability methods for factor-number estimation by exploiting information from the entire spectrum. Johnstone and Onatski \cite{johnstone2020testing} studied the detectability of weak subcritical factors below the phase-transition threshold, where no sample eigenvalue separates from the bulk, and derived asymptotic power envelopes for spike detection. However, their proposal is not directly operational as a practical factor-number estimator.

The original operational procedure of Onatski \cite{onatski2009testing} is not formulated explicitly in terms of the BBP transition. Moreover, when the distinction between strong and weak factors is not sharp, the test may behave as if no additional factors were present. The goal of the present work is to estimate the number of global factors in the vicinity of the BBP transition. We assume a factor model with spiked population structure,
\[
\mathbf{\Sigma}=\operatorname{diag}(\lambda_1,\ldots,\lambda_K,\sigma^2,\ldots,\sigma^2).
\]
The first step is to iteratively identify eigenvalues that deviate from the boundary of the Marčenko--Pastur bulk, under assumptions related to those of Ref.~\cite{kritchman2009non}. To this end, we recalibrate the upper edge of the Marčenko--Pastur support by estimating \(\sigma^2\) from the ratio between the median of the empirical bulk eigenvalues and the median of the Marčenko--Pastur law, following the robust estimation logic used in spiked models \cite{gavish2014optimal}.

In addition, we are interested in detecting global and widespread factors, while avoiding factors localized on only a few assets. For this purpose, we derive asymptotic results for the participation ratio (PR), or equivalently its inverse form, the inverse participation ratio (IPR), which is widely used in the econophysics literature \cite{plerou1999universal}. Previous work derived limits for delocalized graph-Laplacian eigenvectors by averaging the IPR polynomial over a hypersphere \cite{clark2018moments}. Here, we show that weak-factor directions and typical idiosyncratic eigenvectors of the Brown--Harding~(BH) factor model~\cite{brown1989number,harding2008explaining} satisfy the same delocalized benchmark in the high-dimensional regime, whereas the leading coherent common factor satisfies a strong common-loading limit congruent with an economic market factor.

These results motivate a PR criterion for selecting extended factors. Specifically, a candidate factor must first separate spectrally from the residual bulk and then, as a confirmatory criterion, exceed a PR threshold associated with delocalized eigenvectors. The resulting procedure is called the iterative global factor (IGF) algorithm. In Monte Carlo simulations of the BH model, the IGF algorithm recovers the true number of factors near the BBP transition. In this setting, the Onatski test, despite being a state-of-the-art benchmark, fails to recover the true number of factors, while classical criteria based on the Tracy--Widom distribution and the Marčenko--Pastur upper edge require substantially larger dimensions to converge to the correct value.

Finally, the IGF algorithm is applied to the S\&P 500 returns data set studied in Ref.~\cite{molero2023market}. The method detects a larger and more dynamic set of global factors than the Onatski test, suggesting the presence of weak but extensive market components that may lie close to the BBP transition.

Section~2 introduces the RMT background and the main statistical procedures for estimating the number of factors in high-dimensional settings. Section~3 presents the IGF algorithm and the associated PR limits. Section~4 discusses Monte Carlo simulations in a controlled BH factor model setting. Section~5 applies the method to the empirical S\&P 500 data set. Finally, Sec.~6 summarizes the main conclusions and outlines possible directions for future work.

\section{Random matrix theory for factor analysis}

Let $\mathbf{X}$ be a $p\times n$ data matrix whose columns are independent Gaussian $p$-dimensional vectors with zero mean and population covariance matrix $\mathbf{\Sigma}$. The product
\(\mathbf{W}=\mathbf{X}\mathbf{X}^{\top}\) is a Wishart matrix, also known in the physics literature as the real Wishart, or Laguerre orthogonal, ensemble. In mathematical statistics, \(\mathbf{W}\) has a $p$-variate Wishart distribution with \(n\) degrees of freedom, denoted by \(W_p(n,\mathbf{\Sigma})\). The associated sample covariance matrix is
\[
    \mathbf{S}=\frac{1}{n}\mathbf{X}\mathbf{X}^{\top}.
\]
An important RMT result in the context of econophysics is that, when the number of variables (assets) \(p\) and observations (trading days) \(n\) grow without bound while the ratio \(q=p/n\) remains fixed, the eigenvalue distribution of \(\mathbf{S}\) converges to the Marčenko--Pastur law \cite{marvcenko1967distribution},
\begin{equation}
    \rho(\lambda) =
    \frac{\sqrt{(\lambda_{+}-\lambda)(\lambda-\lambda_{-})}}
    {2\pi q \sigma^2\lambda},
    \quad
    \lambda_{\pm} = \sigma^2 (1\pm\sqrt{q})^2,
    \label{MP}
\end{equation}
under the isotropic-noise assumption \(\mathbf{\Sigma}=\sigma^2\mathbf{I}\).

In the early 2000s, Refs.~\cite{laloux1999noise,plerou2002random} proposed a method to filter, or denoise, empirical covariance matrices based on the Marčenko--Pastur law. The idea is to identify as signals the eigenvalues that exceed the upper edge \(\lambda_{+}\) and then reconstruct the covariance matrix accordingly. Although this method was initially designed for noise reduction, it also provides a natural criterion for detecting the number of signals or components in high-dimensional data.

A more formal statistical procedure based on the same central ideas was implemented in Ref.~\cite{molero2023market}. The authors test the null hypothesis
\[
    H_0:\mathbf{\Sigma}=\mathbf{I},
\]
against the alternative hypothesis
\[
    H_a:\mathbf{\Sigma}\neq\mathbf{I},
\]
where \(\mathbf{\Sigma}\) has a more general structure. Under this approach, confidence intervals can be constructed to test deviations from the universal Wishart predictions and applied to empirical samples for arbitrary values of \(p\) and \(n\). These confidence intervals are based on the Tracy--Widom distribution of the largest eigenvalue \(\lambda_1\) of Wishart matrices \cite{johnstone2001distribution}. In particular, the largest eigenvalue of several random-matrix ensembles satisfies the asymptotic relation \cite{tracy1994level}
\begin{equation}
    P\{ n\hat{\lambda}_1 \leq \mu_{np} + \sigma_{np} s \mid H_0\}
    \rightarrow F_{\beta}(s),
    \label{tracy_widom}
\end{equation}
where \(\mu_{np}\) and \(\sigma_{np}\) are centering and scaling parameters that depend on the particular RMT ensemble. This result is known as the Tracy--Widom law. The parameter \(\beta\) takes the values \(1\), \(2\), and \(4\), depending on the data type or symmetry class. The case of interest here is \(\beta=1\), which corresponds to real-valued data with Wishart structure. The convergence rate in Eq.~(\ref{tracy_widom}) is of order \(\mathcal{O}(p^{-1/3})\). However, with modified centering and scaling parameters, the approximation error can be reduced to \(\mathcal{O}(p^{-2/3})\), which makes the approximation useful for relatively small samples \cite{johnstone2008multivariate}.

A state-of-the-art test for determining the number of significant factors in high-dimensional econometric settings was proposed by Onatski \cite{onatski2008tracy,onatski2009testing}. Onatski showed that the distribution of the first \(r\) centered and scaled eigenvalues of a complex Wishart matrix converges weakly to the \(r\)-dimensional joint Tracy--Widom distribution. This result provides the basis for the \(R\) statistic proposed in Ref.~\cite{onatski2009testing} to determine the number of factors in the generalized dynamic factor model (DFM) framework of Ref.~\cite{forni2000generalized}.

The practical procedure can be summarized as follows:

\begin{itemize}
   \item[(i)] Given the data matrix \(\mathbf{X}\), divide it into two time periods of equal size and construct the complex matrix
   \begin{equation}
        \hat{\mathbf{X}}_j
        =
        \mathbf{X}_j
        +
        \mathrm{i}\mathbf{X}_{j+n/2}.
   \end{equation}

   \item[(ii)] Compute the eigenvalues \(\hat{\lambda}_1,\dots,\hat{\lambda}_p\) of
   \begin{equation}
        \frac{2}{n}\hat{\mathbf{X}}\hat{\mathbf{X}}^{\dagger},
   \end{equation}
   where \(\dagger\) denotes conjugate transposition.

   \item[(iii)] Calculate the statistic
   \begin{equation}
        \hat{R}
        =
        \max_{k_0<i\leq k_1}
        \frac{\hat{\lambda}_i-\hat{\lambda}_{i+1}}
        {\hat{\lambda}_{i+1}-\hat{\lambda}_{i+2}} .
   \end{equation}

   \item[(iv)] Reject \(H_0\) according to the decision rule determined by the critical values of the statistic \(\hat{R}\) (see Appendix~\ref{appendix:a}).
\end{itemize}

The proof of the test is based on the spectral separation of the signals associated with the factors and is related to the joint Tracy--Widom distribution for \(\beta=2\) (further details are given in Ref.~\cite{molero2023market}). In practical terms, the procedure for determining the number of factors consists of recursively applying the previous steps and testing a different number of factors at each iteration until the null hypothesis is no longer rejected. Thus, it is assumed, a priori, that the true number of factors \(k\) lies between \(k_1\) and \(k_2\). One first tests \(H_0:k=k_1\) against \(H_1:k_1<k\leq k_2\) at significance level \(\alpha\). If \(H_0\) is not rejected, the estimated number of factors is \(k_1\), and the algorithm stops. If \(H_0\) is rejected, one tests \(H_0:k=k_1+1\) against \(H_1:k_1+1<k\leq k_2\). The procedure is repeated until \(H_0\) is not rejected, and the corresponding number of factors is taken as the estimate for the associated model.

\section{Iterative global factor estimator}

The proposed IGF algorithm is based on four main elements. First, we assume a static approximate factor model with dense random loadings and isotropic idiosyncratic noise. We refer to this model as the Brown--Harding (BH) factor model because it was introduced by Brown~\cite{brown1989number}, and the corresponding asymptotic correction for the distribution of sample eigenvalues was later discussed by Harding~\cite{harding2008explaining}. Second, we show that, under suitable parameter conditions, the spike-like structure of the BH model is approximately preserved after transforming the covariance matrix into a correlation matrix. Third, we introduce an iterative criterion for detecting signals by recalibrating the effective noise level at each step. Fourth, we incorporate an eigenvector delocalization criterion based on an asymptotic PR threshold, derived in detail in Appendix~\ref{appendix:c}. Each of these elements is described below, culminating in the presentation of the IGF algorithm.

\subsection{Brown-Harding factor model}
\label{sec:bh_model}

Consider the Brown--Harding (BH) factor model with dense random loadings and isotropic idiosyncratic noise,
\begin{equation}
    \mathbf{\Sigma}_{BH}
    =
    \sigma_f^2 \mathbf{L}\mathbf{L}^{\top}
    +
    \sigma_e^2\mathbf{I},
    \label{model_p1}
\end{equation}
where the covariance matrix of the factors is assumed to be
\(\mathbf{\Sigma}_f=\sigma_f^2\mathbf{I}\), so that the factors are uncorrelated and have the same variance. The loading matrix \(\mathbf{L}\in\mathbb{R}^{p\times K}\) generates \(K\) dominant directions. Thus, \(\sigma_f^2\mathbf{L}\mathbf{L}^{\top}\) represents the spike structure, whereas \(\sigma_e^2\mathbf{I}\) represents isotropic idiosyncratic noise. We consider nearly homogeneous i.i.d. Gaussian factor loadings,
\begin{equation}
    L_{ik}
    =
    b+\eta_{ik},
    \quad
    \eta_{ik}\sim \mathcal{N}(0,\sigma_b^2),
    \quad
    i=1,\dots,p,\quad k=1,\dots,K.
    \label{model_p2}
\end{equation}

In the eigenbasis of \(\mathbf{L}\mathbf{L}^{\top}\), the population covariance matrix \(\mathbf{\Sigma}_{BH}\) has the diagonal spiked form \cite{Iain2006}
\begin{equation}
    \mathbf{\Sigma}_{BH}
    =
    \operatorname{diag}
    (\lambda_1,\dots,\lambda_K,\sigma_e^2,\dots,\sigma_e^2),
    \label{model_p3}
\end{equation}
where the population eigenvalues are
\begin{equation}
    \lambda_j
    =
    p\sigma_f^2(\sigma_b^2+Kb^2\delta_{j1})
    +
    \sigma_e^2,
    \quad
    j=1,\dots,K.
\end{equation}
Here, \(\delta_{j1}\) denotes the Kronecker delta. These models are called spiked models because they contain a finite number of isolated population eigenvalues embedded in a noise background of scale \(\sigma_e^2\) \cite{johnstone2001distribution}.

In high dimensions, the sample eigenvalues associated with idiosyncratic noise are spread according to the Marčenko--Pastur distribution with scale parameter \(\sigma_e^2\). In contrast, the sample eigenvalue associated with the strongest signal is upward biased, whereas the remaining \(K-1\) weaker signals are subject to a phase-transition phenomenon \cite{baik2006eigenvalues}. These effects are characterized by Harding's sample corrections, which describe the almost-sure convergence of the sample eigenvalues to the following deterministic limits \cite{harding2008explaining}:
\begin{equation}
\hat{\lambda}_i^H \rightarrow
\begin{cases}
\left[p\sigma_f^2(\sigma_b^2+Kb^2)+\sigma_e^2\right]
\left[
1+\dfrac{1}{n}
\dfrac{\sigma_e^2}{\sigma_f^2(\sigma_b^2+Kb^2)}
\right],
& i=1, \\[1.0em]

\left(p\sigma_f^2\sigma_b^2+\sigma_e^2\right)
\left[
1+\dfrac{1}{n}
\dfrac{\sigma_e^2}{\sigma_f^2\sigma_b^2}
\right],
& i=2,\dots,K,\quad
p\geq p_c, \\[1.2em]

\lambda_+,
& i=2,\dots,K,\quad
p<p_c, \\[1.2em]

\lambda_+,
& i=K+1,\dots,p .
\end{cases}
\label{model_p4}
\end{equation}
where
\begin{equation}
    p_c
    :=
    \dfrac{1}{n}
    \left(
    \dfrac{\sigma_e^2}{\sigma_f^2\sigma_b^2}
    \right)^2
\end{equation}
denotes the critical threshold separating the two regimes of the phase transition.

\subsection{Correlation structure of Brown-Harding factor model}

The object of interest is the correlation matrix, since it removes differences in scale across assets. To obtain the correlation matrix from the covariance matrix, we use the standard transformation
\begin{equation}
    \mathbf{C}
    =
    \mathbf{D}^{-1/2}
    \boldsymbol{\Sigma}
    \mathbf{D}^{-1/2},
    \quad
    \mathbf{D}
    =
    \operatorname{diag}(\boldsymbol{\Sigma}).
\end{equation}

For the BH factor model, this gives
\begin{align}
    \mathbf{C}_{BH}
    &=
    \mathbf{D}^{-1/2}
    \left(
        \sigma_f^2 \mathbf{L}\mathbf{L}^{\top}
        +
        \sigma_e^2 \mathbf{I}
    \right)
    \mathbf{D}^{-1/2}
    \nonumber \\
    &=
    \sigma_f^2
    \mathbf{D}^{-1/2}
    \mathbf{L}\mathbf{L}^{\top}
    \mathbf{D}^{-1/2}
    +
    \sigma_e^2\mathbf{D}^{-1}.
    \label{eq:correlation_decomposition}
\end{align}
Equivalently,
\begin{equation}
    \mathbf{C}_{BH}
    =
    \sigma_f^2
    \widetilde{\mathbf{L}}
    \widetilde{\mathbf{L}}^{\top}
    +
    \sigma_e^2\mathbf{D}^{-1},
    \quad
    \widetilde{\mathbf{L}}
    =
    \mathbf{D}^{-1/2}\mathbf{L}.
    \label{eq:normalized_loadings}
\end{equation}

For each asset, the corresponding diagonal entry is
\begin{equation}
    D_{ii}
    =
    \sigma_e^2
    +
    \sigma_f^2 \ell_i^2,
    \quad
    \ell_i^2
    =
    \sum_{\alpha=1}^{K}L_{i\alpha}^2 .
    \label{eq:Dii}
\end{equation}
Using the distribution of \(L_{i\alpha}\) in Eq.~\eqref{model_p2}, the mean population variance scale is
\begin{equation}
    \langle D_{ii} \rangle
    =
    \sigma_e^2
    +
    \sigma_f^2 K(b^2+\sigma_b^2)
    \equiv
    d_0 .
    \label{eq:d0}
\end{equation}
Thus, when the diagonal entries are close to their typical scale, the BH factor model can be represented approximately in correlation form as
\begin{equation}
    \mathbf{C}_{BH}
    \approx
    \sigma_f^2
    \widetilde{\mathbf{L}}
    \widetilde{\mathbf{L}}^{\top}
    +
    \sigma_0^2\mathbf{I},
    \quad
    \sigma_0^2
    =
    \frac{\sigma_e^2}{d_0}.
    \label{model_p5}
\end{equation}
The diagonal approximation of the rescaled idiosyncratic noise is valid when the coefficient of variation of \(D_{ii}\) satisfies \(CV_D\ll 1\). In particular, the coefficient of variation of the diagonal entries is given by (see Appendix~\ref{appendix:b})
\begin{equation}
    CV_D
    =
    \frac{
    \sigma_f^2
    \sqrt{
    K(2\sigma_b^4+4b^2\sigma_b^2)
    }
    }{
    \sigma_e^2+\sigma_f^2K(b^2+\sigma_b^2)
    }.
    \label{eq:cvD}
\end{equation}

For the parameters used in Sec.~\ref{4.Simulations}, we obtain \(CV_D\approx 0.0352\). Thus, the diagonal entries have a dispersion of approximately \(3.52\%\) relative to the typical scale \(d_0\).

\subsection{Iterative recalibration}

The next element consists of estimating the effective noise level \(\hat{\sigma}_0^2\) of the factor model from the empirical correlation matrix. For this purpose, we use a procedure inspired by Ref.~\cite{gavish2014optimal}, where robust estimation based on the median is proposed in the context of the singular values of a low-rank rectangular matrix contaminated with noise. The intuition is that the spikes separate from the bulk, while the median is not strongly affected by a finite number of isolated eigenvalues. We follow the same logic. The null model for the observed correlation matrix is \(\mathbf{C}_0=\mathbf{I}\), whereas, if the data are generated by a static approximate factor model with dense random loadings and isotropic idiosyncratic noise, the alternative model is \(\mathbf{C}_a=\mathbf{C}_{BH}\).

The factor model in correlation form still has a spiked structure. Therefore, the empirical bulk eigenvalues of \(\mathbf{C}_{BH}\) can be approximated by
\begin{equation}
    \lambda_{\mathrm{bulk}}(\mathbf{C}_{BH})
    \approx
    \sigma_0^2 \lambda(\mathbf{I}),
    \quad
    \lambda(\mathbf{I})\sim MP(q,1),
\end{equation}
where \(MP(q,1)\) denotes the unit-scale Marčenko--Pastur law with aspect ratio \(q\) (see Eq.~\eqref{MP}). In particular, the argument of Ref.~\cite{gavish2014optimal} motivates the following asymptotic estimator of \(\sigma_0^2\) from the sample eigenvalues:
\begin{equation}
    \hat{\sigma}_0^2
    =
    \frac{
    \operatorname{median}
    \left[
    \tilde{\lambda}_{\mathrm{bulk}}(\mathbf{C}_{BH})
    \right]
    }{
    m_{\mathrm{MP}}(q;1)
    },
\end{equation}
where the numerator is the median of the empirical bulk eigenvalues and \(m_{\mathrm{MP}}(q;1)\) is the median of the unit-scale Marčenko--Pastur distribution.. This is the central element of the IGF estimator. We iteratively test the \(k\)th component by excluding the previous \(k-1\) detected signals and estimating the effective variance from the remaining bulk.

Let \(\tilde{\lambda}_1\geq \tilde{\lambda}_2\geq \cdots \geq \tilde{\lambda}_p\) be the empirical eigenvalues of the sample correlation matrix. Then, the effective variance at step \(k\) is given by
\begin{equation}
    \hat{\sigma}_0^2(k)
    =
    \frac{
    \operatorname{median}
    \left(
    \tilde{\lambda}_{k},
    \tilde{\lambda}_{k+1},
    \dots,
    \tilde{\lambda}_p
    \right)
    }{
    m_{\mathrm{MP}}(q;1)
    },
    \quad
    q_k=\frac{p-k+1}{n},
\end{equation}
Thus, we construct an adaptive upper edge of the noise Marčenko--Pastur distribution for the \(k\)th signal,
\begin{equation}
    \widehat{\lambda}_{+,k}
    =
    \widehat{\sigma}_{0}^{2}(k)
    \left(1+\sqrt{q_k}\right)^2.
\end{equation}
The spectral decision rule for considering \(\tilde{\lambda}_k\) a candidate factor signal is
\begin{equation}
    \tilde{\lambda}_k>\widehat{\lambda}_{+,k}.
\end{equation}

This sequential estimation is related to Lemma 1 of Ref.~\cite{kritchman2009non}, where it is shown that, after removing the spike eigenvalues, the next largest eigenvalue follows a Tracy--Widom distribution. This observation is especially important near the BBP transition. In that regime, a weak factor can produce a sample eigenvalue only slightly above the noise edge. The recalibration of \(\widehat{\sigma}_{0}^{2}(k)\) uses the residual spectrum to adapt the noise edge at each step, thereby improving the resolution of weak but spectrally separated factors.

In practice, we use the asymptotic approximation \(q=p/n\) instead of \(q_k\), because the number of detected signals is expected to be small compared with the number of assets. The median of the Marčenko--Pastur distribution was computed numerically using the composite trapezoidal rule, as implemented in NumPy's \texttt{trapz} function \cite{harris2020array}.

\subsection{Participation-ratio limits}

The last element is an eigenvector-level confirmation criterion designed to filter out signals that are not sufficiently extended across the asset universe. Consider an arbitrary normalized eigenvector \(u_k\), \(k=1,\dots,p\), of the correlation matrix. The PR of \(u_k\) is defined as
\begin{equation}
\mathrm{PR}(u_k)
=
\frac{1}{\sum_{i=1}^{p}u_{ik}^4}.
\end{equation}

The PR lies between \(1\) and \(p\). In a fully localized direction, where only one component contributes, the PR reaches its minimum value \(1\). In a fully extended and uniformly distributed direction, the PR reaches its maximum value \(p\). In financial terms, \(\mathrm{PR}=p\) corresponds to a maximally diversified factor, while \(\mathrm{PR}=1\) corresponds to a completely localized portfolio concentrated in a single asset. If only \(m\) assets contribute approximately uniformly, one expects \(\mathrm{PR}\approx m\).

A contribution of this work is to derive asymptotic PR benchmarks for the BH factor model and use them as an operational eigenvector-level criterion. In the strong common-loading regime and for large \(p\), the leading coherent direction spreads approximately uniformly over all assets, so that
\[
\frac{\mathrm{PR}(u_1)}{p}\to 1 .
\]
By contrast, weak-factor directions and typical idiosyncratic sample eigenvectors behave asymptotically as random real-delocalized directions and satisfy (see Appendix~\ref{appendix:c})
\begin{equation}
\frac{\mathrm{PR}(u_k)}{p}\to \frac{1}{3}, \quad k=2,\dots,p .
\label{extensive_ipr}
\end{equation}
This result connects the BH factor model with the real-delocalized eigenvector benchmark. We therefore use the normalized PR as an additional criterion to distinguish localized eigenvectors from extended global-factor candidates.

\subsection{The IGF algorithm}

The IGF procedure can be summarized as follows:

\begin{enumerate}
    \item Compute the empirical correlation matrix \(\tilde{\mathbf C}\) and its
    eigendecomposition
    \[
        \tilde{\mathbf C}\tilde{\mathbf u}_k
        =
        \tilde\lambda_k \tilde{\mathbf u}_k,
        \qquad
        \tilde\lambda_1\geq \tilde\lambda_2\geq \cdots \geq \tilde\lambda_p .
    \]

    \item Set \(\widehat r_\tau=0\). For \(k=1,\dots,k_{\max}\), estimate the
    residual noise scale from the remaining spectrum as
    \[
        \widehat{\sigma}_0^2(k)
        =
        \frac{
        \operatorname{median}
        \{\tilde\lambda_k,\tilde\lambda_{k+1},\ldots,\tilde\lambda_p\}
        }{
        m_{\mathrm{MP}}(q;1)
        },
        \qquad
        q=\frac{p}{n},
    \]

    \item Compute the corresponding upper edge,
    \[
        \widehat{\lambda}_{+,k}
        =
        \widehat{\sigma}_0^2(k)(1+\sqrt q)^2 .
    \]

    \item If
    \[
        \tilde\lambda_k \leq \widehat{\lambda}_{+,k},
    \]
    stop the sequential search.

    \item Otherwise, compute the normalized participation ratio
    \[
        \frac{\mathrm{PR}(\tilde{\mathbf u}_k)}{p}
        =
        \frac{1}{p\sum_{i=1}^p \tilde u_{ik}^4}.
    \]
    If
    \[
        \frac{\mathrm{PR}(\tilde{\mathbf u}_k)}{p}\geq \tau,
    \]
    update
    \[
        \widehat r_\tau \leftarrow \widehat r_\tau+1.
    \]
    If the PR condition is not satisfied, do not count the component, but
    continue with the next value of \(k\).

    \item Return \(\widehat r_\tau\).
\end{enumerate}

The sequential search stops when the candidate eigenvalue no longer exceeds
the iteratively estimated Marčenko--Pastur upper edge, or when \(k_{\max}\)
is reached. Since the \(\mathrm{PR}/p\) criterion is applied locally to each
spectrally separated component and is not necessarily monotone in the
eigenvalue rank, the algorithm can accept noncontiguous eigenvalue indices
as global factors.

\section{Simulations}
\label{4.Simulations}

\begin{figure}[htbp]
    \centering
    \begin{subfigure}[b]{0.7\textwidth}
        \centering
        \includegraphics[width=\textwidth]{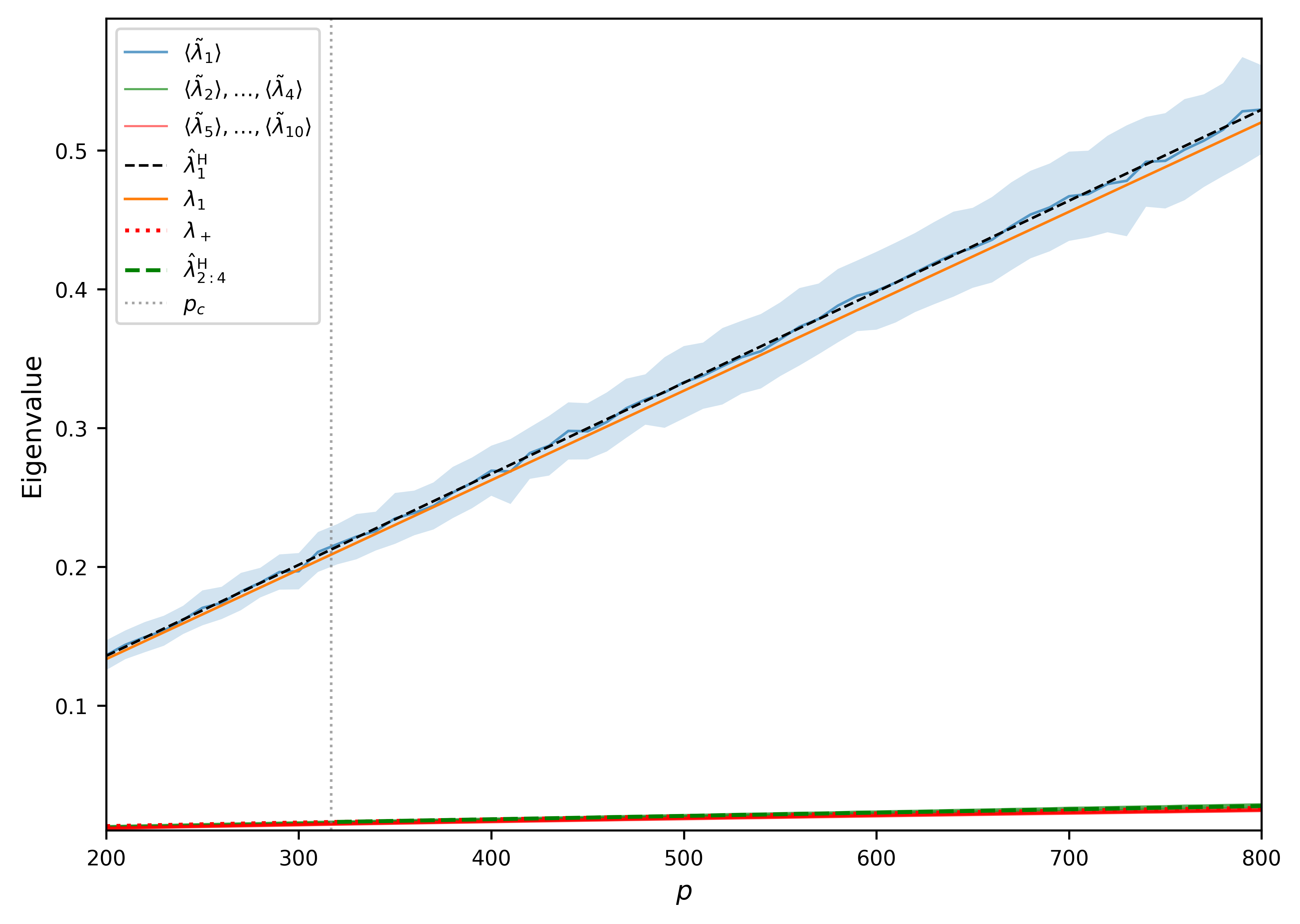}
        \caption{}
    \end{subfigure}\\
    \begin{subfigure}[b]{0.7\textwidth}
        \centering
        \includegraphics[width=\textwidth]{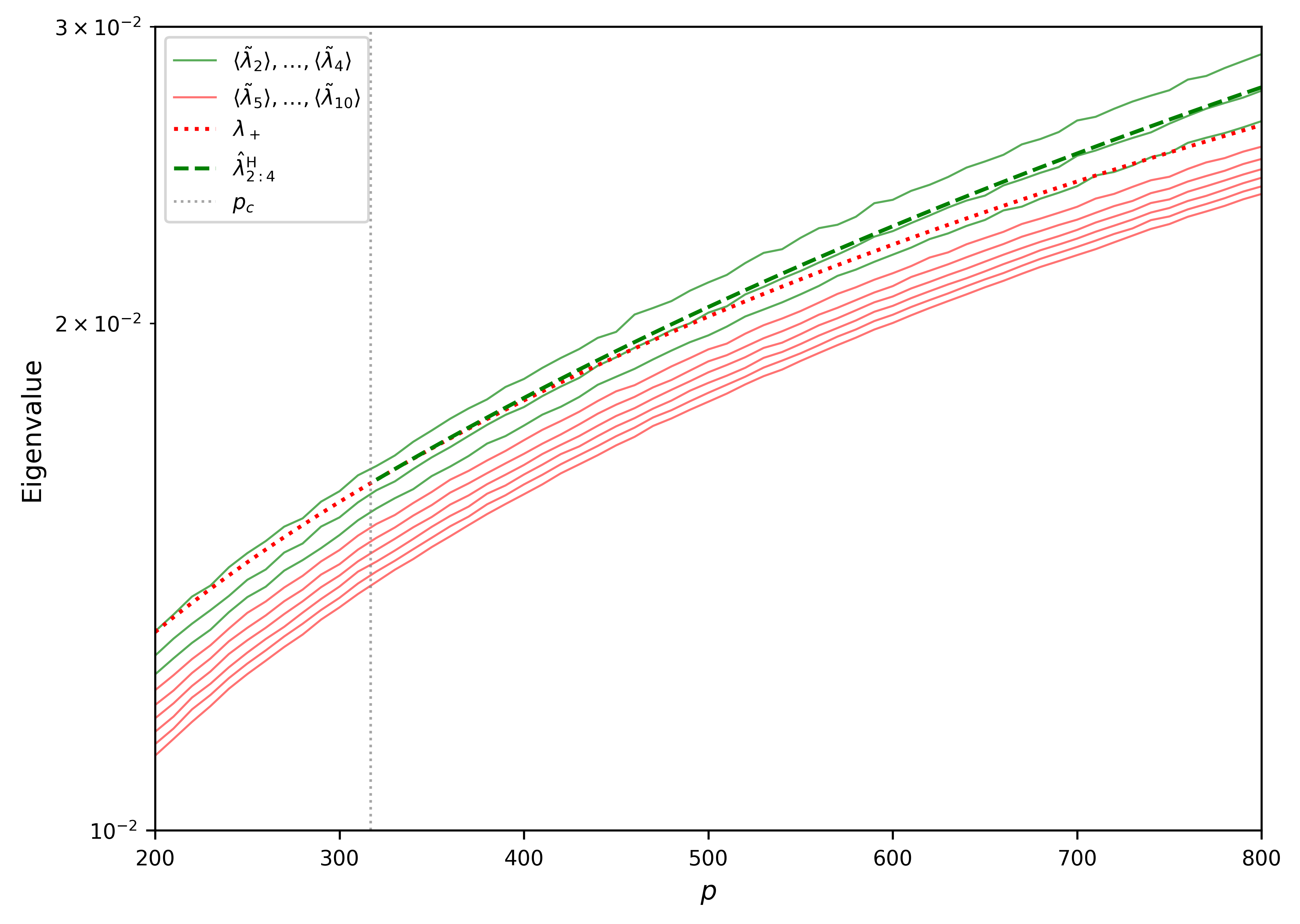}
        \caption{}
    \end{subfigure}
    \caption{Dimension-scaling analysis of the BH factor model. Panel (a) shows the mean of the first ten sample eigenvalues over 100 Monte Carlo replications for $p=200,\ldots,800$, with $\Delta p=10$ and $n=400$. The parameters are $\sigma_b^2=0.08$, $\sigma_f^2=0.000158$, $\sigma_e^2=0.0045$, and $b=1$. Shading denotes one standard deviation of the largest sample eigenvalue. The plot also shows the largest population eigenvalue, Harding's corrections for the first four sample eigenvalues, the Marčenko--Pastur upper edge, and the critical point $p_c$. Panel (b) shows the weak-factor regime on a semilogarithmic scale.}
    \label{fig:brown_dimension_scaling}
\end{figure}

Figure~\ref{fig:brown_dimension_scaling}(a) shows the behavior of the first ten eigenvalues of the BH factor model described by Eqs.~\eqref{model_p1}--\eqref{model_p3}, obtained from Monte Carlo simulations with 100 replications. Here, $p$ is varied over the grid $[200,800]$ with $\Delta p=10$, while the sample size $n=400$ and the model parameters $\sigma_b^2=0.08$, $\sigma_f^2=0.000158$, $\sigma_e^2=0.0045$, and $b=1$ are held fixed.

The mean largest sample eigenvalue $\langle\tilde{\lambda}_1\rangle$ shows a positive bias relative to the largest population eigenvalue $\lambda_1$, but this bias is accurately described by Harding's deterministic limit $\hat{\lambda}_1^H$ in Eq.~\eqref{model_p4}. The shaded region represents one standard deviation of the largest sample eigenvalue. The mean sample eigenvalues associated with the weak factors, $\langle \tilde{\lambda}_2 \rangle,\dots,\langle \tilde{\lambda}_4 \rangle$, are shown in green, while a representative set of idiosyncratic mean sample eigenvalues, $\langle \tilde{\lambda}_5 \rangle,\dots,\langle \tilde{\lambda}_{10} \rangle$, is shown in red.

A semilogarithmic representation of the weak-factor regime is shown in Fig.~\ref{fig:brown_dimension_scaling}(b). Before the critical dimension $p_c$, the weak and idiosyncratic eigenvalues remain close to the upper edge $\lambda_+$. For $p>p_c$, the weak signals begin to separate from the bulk and approach the deterministic limits $\hat{\lambda}_{2:4}^H$ given by the second row of Eq.~\eqref{model_p4}. For the parameters used here, $p_c\approx 317$.

% Harding correction

\begin{figure}[htbp]
    \centering
    \includegraphics[width=0.7\textwidth]{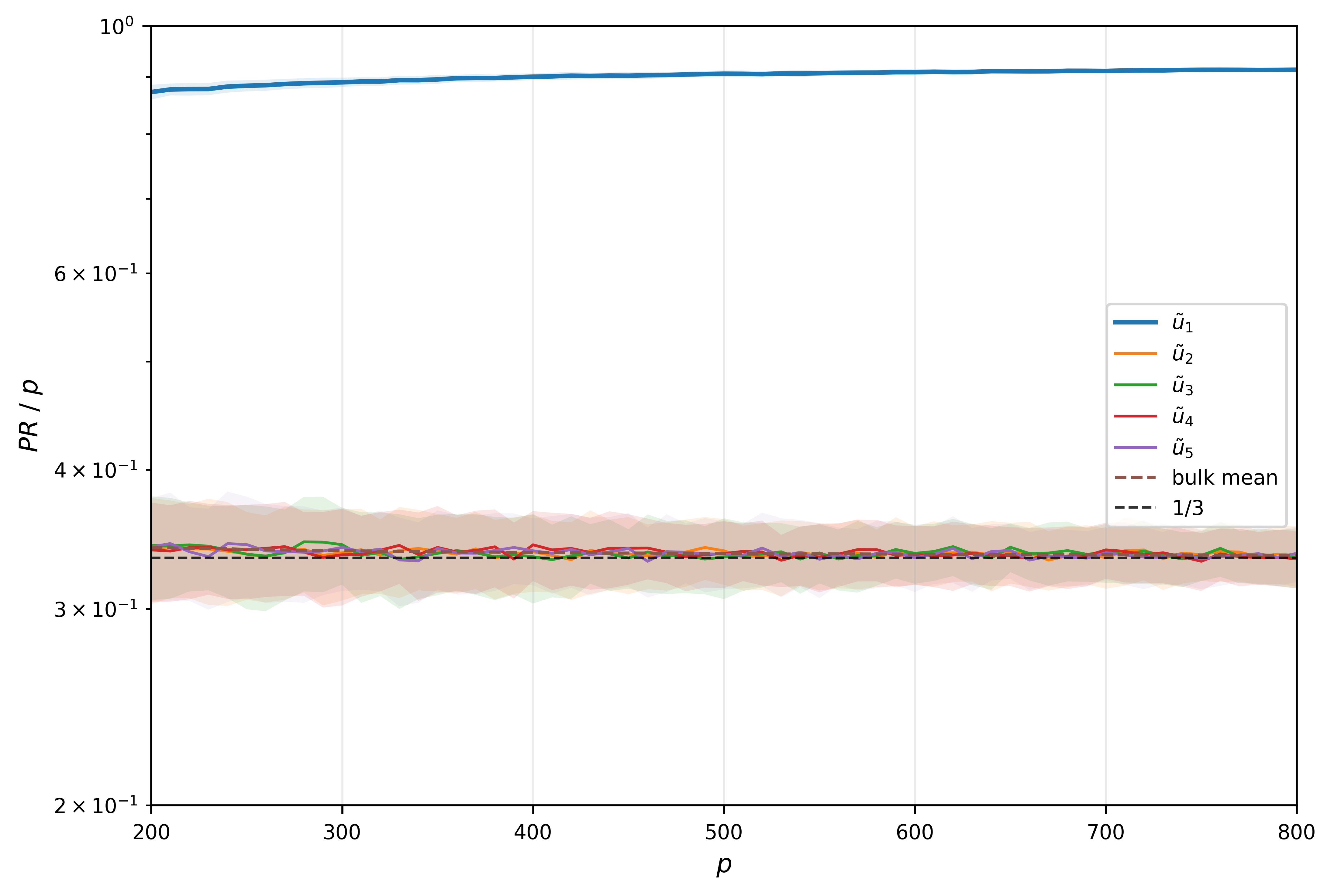}
    \caption{Dimension-scaling behavior of the normalized participation ratio $\mathrm{PR}/p$ in the BH factor model with fixed $q=1/2$. The leading eigenvector remains close to the fully extended limit, while the weak-factor directions fluctuate around the real-delocalized benchmark $1/3$. Shaded regions indicate one standard deviation across 100 Monte Carlo replications.}
    \label{fig:brown_dimension_pr}
\end{figure}

% PR as a function of p (q=1/2)
Figure~\ref{fig:brown_dimension_pr} shows the behavior of $\mathrm{PR}/p$ as the dimension $p$ increases, under the same configuration as Fig.~\ref{fig:brown_dimension_scaling}, except that here $q=1/2$ is fixed. The mean sample eigenvector associated with the largest eigenvalue remains close to 1, indicating a highly diversified direction. By contrast, the mean sample eigenvectors $\tilde{\mathbf{u}}_2$, $\tilde{\mathbf{u}}_3$, and $\tilde{\mathbf{u}}_4$, associated with the weak factors, fluctuate around the asymptotic limit for real-delocalized directions given in Eq.~\eqref{extensive_ipr}. In both cases, the shaded regions represent one standard deviation across Monte Carlo replications. The small deviations from the $1/3$ limit can be attributed to the isotropic approximation $\mathbf{C}\approx d_0^{-1}\mathbf{\Sigma}$ (see Appendix~\ref{appendix:c4}), which is valid when the diagonal heterogeneity is small. In the present setting, the diagonal entries have a dispersion of approximately $3.52\%$ relative to $d_0$, which can account for small corrections around the asymptotic limit.

\begin{figure}[htbp]
    \centering
    \includegraphics[width=0.7\textwidth]{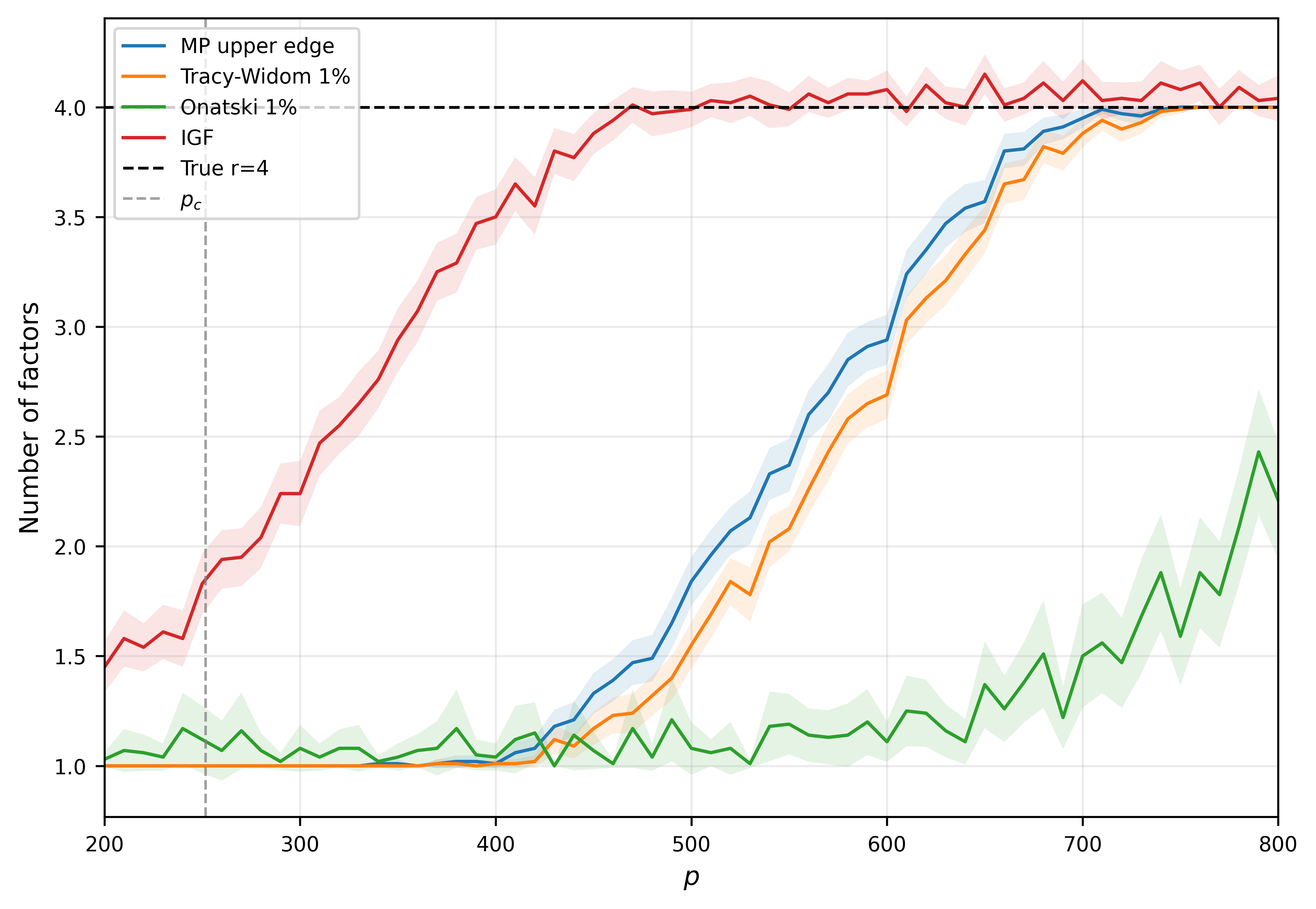}
    \caption{Dimension-scaling behavior of the estimated number of factors at fixed $q=1/2$. The IGF algorithm uses $\tau=0.3$ and $k_{\max}=15$. The figure also shows the Onatski test, the Tracy--Widom criterion at the 1\% level, and the Marčenko--Pastur upper-edge criterion. The vertical dotted line denotes the BBP critical point $p_c\approx252$. Shaded bands represent 95\% Monte Carlo confidence intervals for the mean count across 100 replications.}
    \label{fig:brown_dimension_count}
\end{figure}

% Factor count as a function of p (q=1/2)
Figure~\ref{fig:brown_dimension_count} shows the number of detected factors as the number of assets increases, with fixed $q=1/2$ as in Fig.~\ref{fig:brown_dimension_pr}.
The shaded bands denote 95\% Monte Carlo confidence intervals for the mean factor count across independent replications. Following the setting of Ref.~\cite{molero2023market}, the number of significant factors for the Onatski test was determined at $\alpha\in\{0.01,0.05,0.1\}$. Moreover, an upper bound $k_2=8$ was considered, so the alternative hypothesis was tested only up to that limit, which is sufficient for the present comparison. Here, the IGF algorithm is implemented with $\tau=0.3$, slightly below the asymptotic benchmark for real-delocalized eigenvectors. The heuristic choice of this threshold is discussed in the empirical setting of Sec.~\ref{empirical_data}. The search cutoff is set to $k_{\max}=15$ as an operational upper bound, although this choice does not materially affect the results.

The IGF algorithm recovers the true number of factors at approximately $p=400$, whereas the competing eigenvalue-only criteria require substantially larger dimensions. In particular, the Onatski test does not recover the true number of factors even at $p=800$.
Another point to highlight is that the IGF algorithm detects, on average, more than one factor below the BBP critical point $p_c\approx252$. 

\section{Empirical data}
\label{empirical_data}

The data set consists of companies listed in the S\&P 500. Closing prices were collected for companies with less than 5\% missing observations, and logarithmic returns were then computed. After this filtering step, $p=417$ companies were retained, with missing observations filled by linear interpolation. The study period extends from January 2005 to December 2022.

We set $q=1/2$ to evaluate the explanatory power of the algorithm in the same high-dimensional regime considered in the simulations. This choice gives $n=834$ observations per moving window. In practice, the analyzed period extends until 2022-12-07 and contains 185 moving windows. Successive windows are shifted by $\Delta n=20$ trading days, corresponding roughly to one calendar month. The data set and moving-window construction follow the same empirical setting, allowing for a direct comparison with Ref.~\cite{molero2023market}.

\begin{figure}[htbp]
    \centering
    \includegraphics[width=0.7\textwidth]{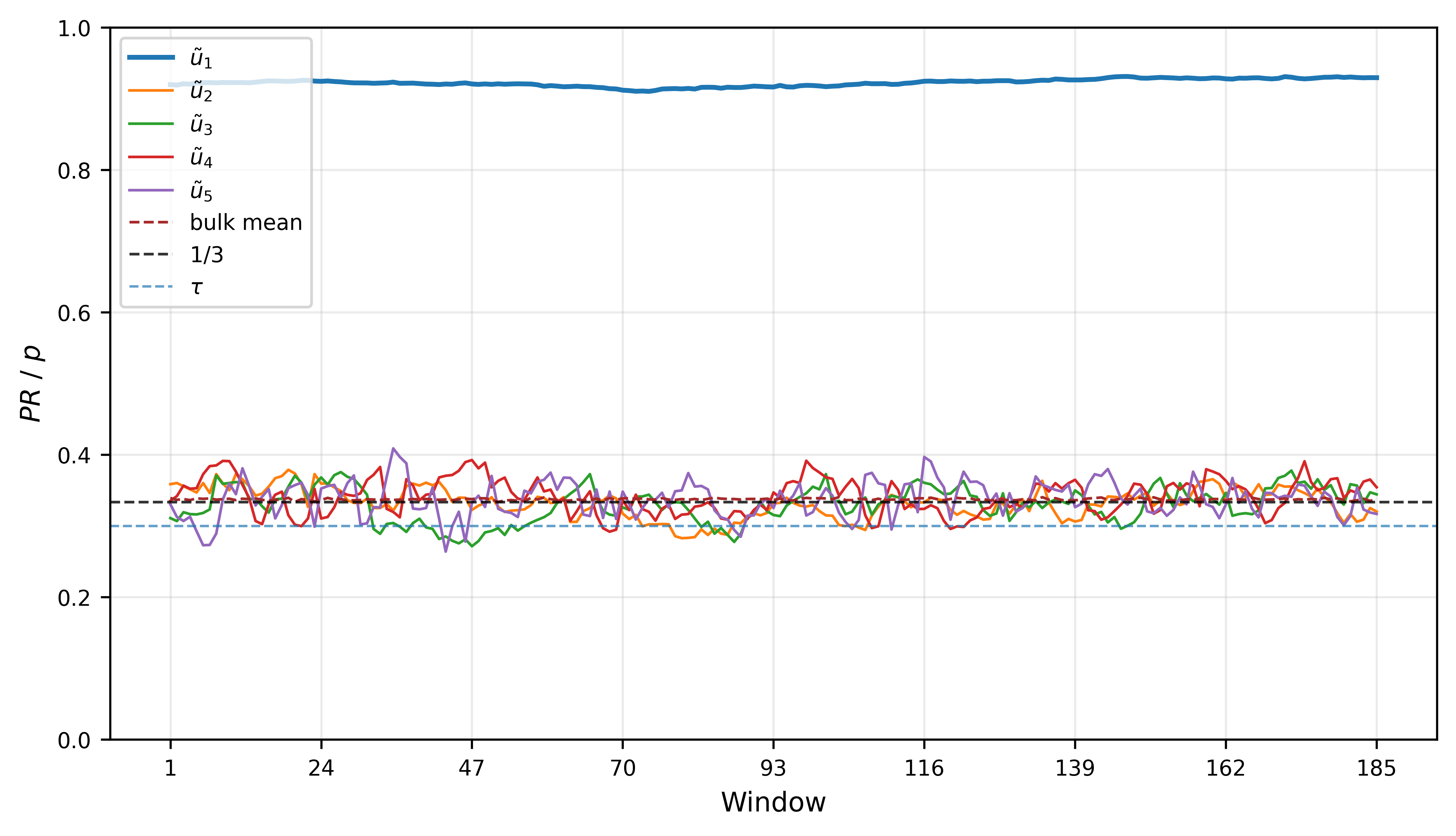}
    \caption{Moving-window mean $\mathrm{PR}/p$ profile for the first five sample eigenvectors in the synthetic BH setting calibrated to the empirical dimensions. The window size is fixed at $p=417$ with $q=1/2$. Dashed lines indicate the bulk mean, the $1/3$ real-delocalized benchmark, and the working threshold $\tau=0.3$, which acts as an effective cutoff for weak-factor participation.}
    \label{fig:brown_windows_pr_profile}
\end{figure}

Before applying the IGF algorithm to real data, the threshold for $\tau$ must be calibrated. To this end, we conduct a controlled experiment that mimics the empirical moving-window setting using the BH factor model as a data-generating process. Figure~\ref{fig:brown_windows_pr_profile} shows the mean $\mathrm{PR}/p$ profile, in the same spirit as Fig.~\ref{fig:brown_dimension_pr}, but using the empirical dimensions in a moving-window scenario. 
Notice that the effective threshold separating the mean weak-factor directions is approximately $\tau=0.3$.

\begin{figure}[htbp]
    \centering
    \includegraphics[width=0.7\textwidth]{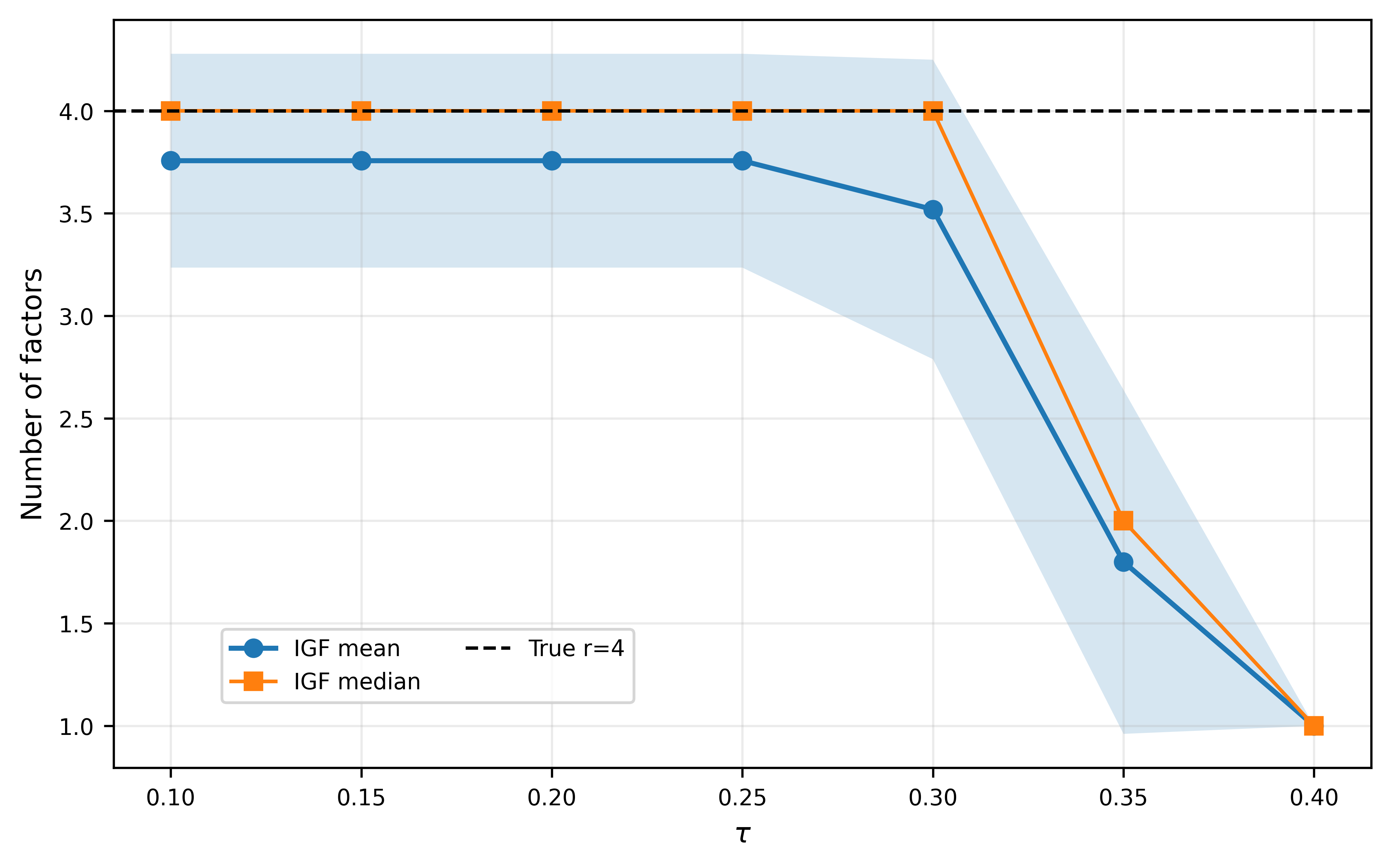}
    \caption{Sensitivity of the IGF factor count to the participation threshold $\tau$ across the $m=185$ synthetic BH moving windows calibrated to the empirical dimensions, with $p=417$ and $q=1/2$.}
    \label{fig:brown_windows_sensitivity}
\end{figure}

Figure~\ref{fig:brown_windows_sensitivity} shows the number of factors detected by the IGF algorithm as the threshold $\tau$ varies over the grid $\{0.1,0.15,0.2,0.25,0.3,0.35,0.4\}$. The median count across the $m=185$ synthetic moving windows recovers the true number of factors at $\tau\leq0.3$. Above this value, the algorithm loses power, and the detected number of factors decreases toward one, corresponding to the detection of only the market factor.

\begin{figure}[htbp]
    \centering
    \includegraphics[width=0.7\textwidth]{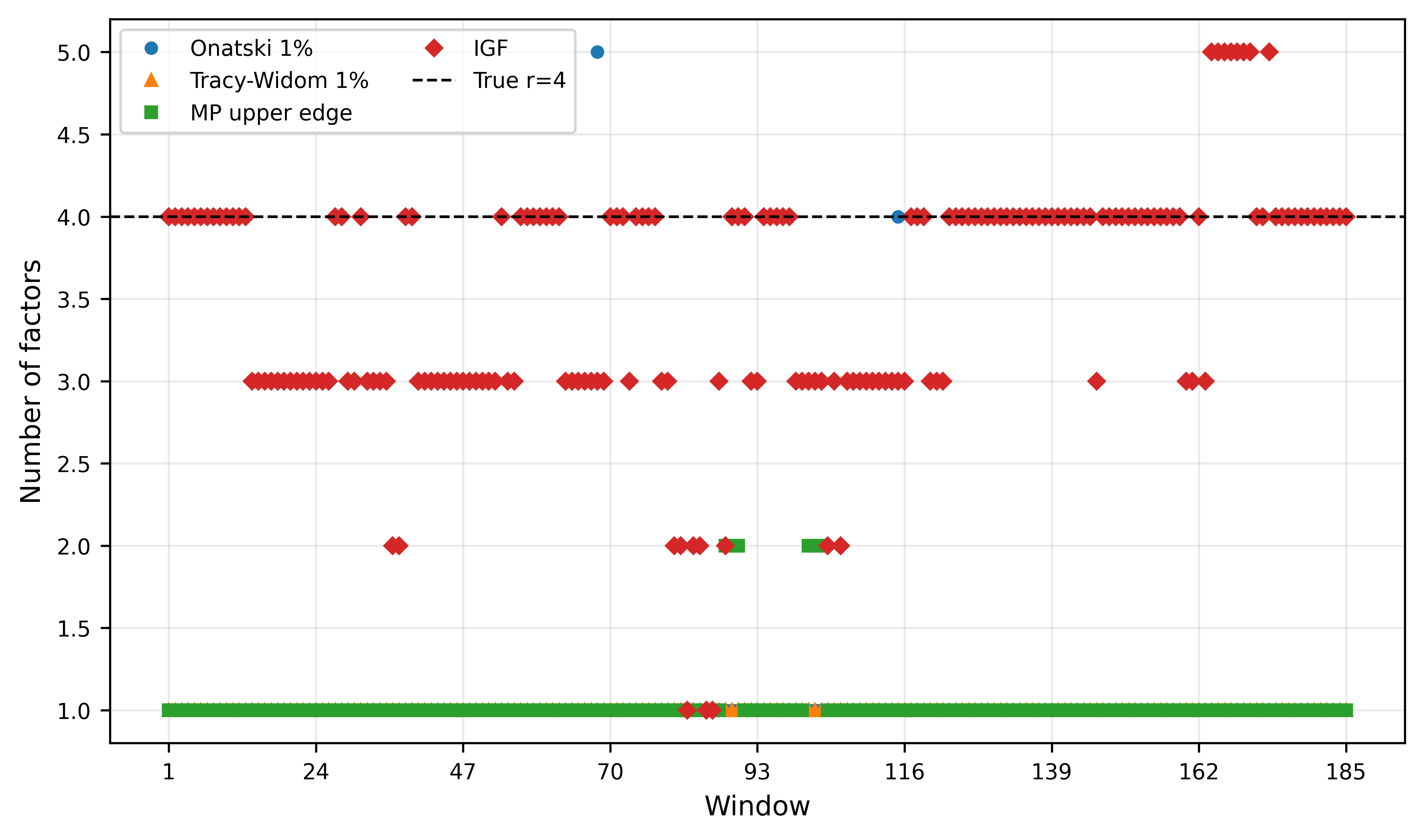}
    \caption{Detected number of factors across moving windows in the synthetic BH model setting. The IGF algorithm uses $\tau=0.3$ and $k_{\max}=15$. The figure also shows the Onatski test, the Tracy--Widom criterion at the 1\% level, and the Marčenko--Pastur upper-edge criterion, using $q=1/2$.}
    \label{fig:brown_windows_counts}
\end{figure}

Figure~\ref{fig:brown_windows_counts} shows the detected number of factors in the synthetic moving-window scenario of the Brown--Harding model. The calibrated threshold $\tau=0.3$ is used for the IGF algorithm. The number of factors fluctuates around 4, as expected. By contrast, the Onatski test does not recover the true number in any window and remains at the one-factor detection level.

% Empirical analysis

\begin{figure}[htbp]
    \centering
    \includegraphics[width=0.7\textwidth]{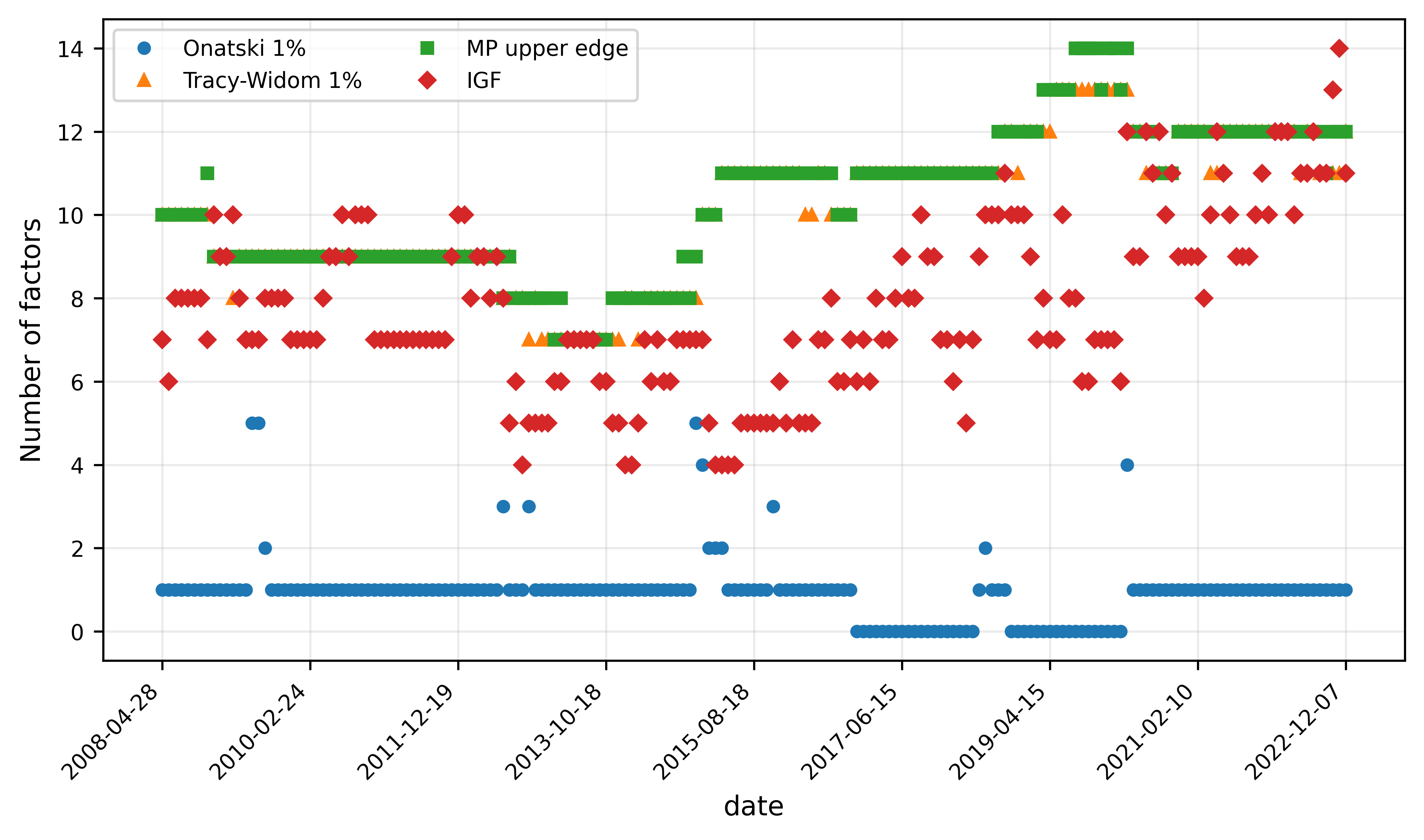}
    \caption{Detected number of factors across moving windows in the empirical S\&P 500 data set. The IGF algorithm uses $\tau=0.3$ and $k_{\max}=15$. The figure also shows the Onatski test, the Tracy--Widom criterion at the 1\% level, and the Marčenko--Pastur upper-edge criterion, using $q=1/2$.}
    \label{fig:empirical_windows_counts}
\end{figure}

Based on the BH moving-window simulation calibrated to the empirical setting, $\tau=0.3$ provides a conservative operational threshold for detecting extensive factors. Accordingly, we adopt the same threshold in the empirical analysis. Figure~\ref{fig:empirical_windows_counts} shows the number of factors detected by each estimator on the empirical S\&P 500 data set. The Onatski test detects only one factor in most windows, as previously reported in Ref.~\cite{molero2023market}. By contrast, the IGF algorithm exhibits richer dynamics, with the detected number of factors fluctuating roughly between 4 and 14. Its behavior is closely related to the Tracy--Widom and Marčenko--Pastur upper-edge criteria, but IGF is more conservative because it restricts detection to delocalized factors and updates the upper edge at each iteration.

\begin{figure}[htbp]
    \centering
    \includegraphics[width=0.7\textwidth]{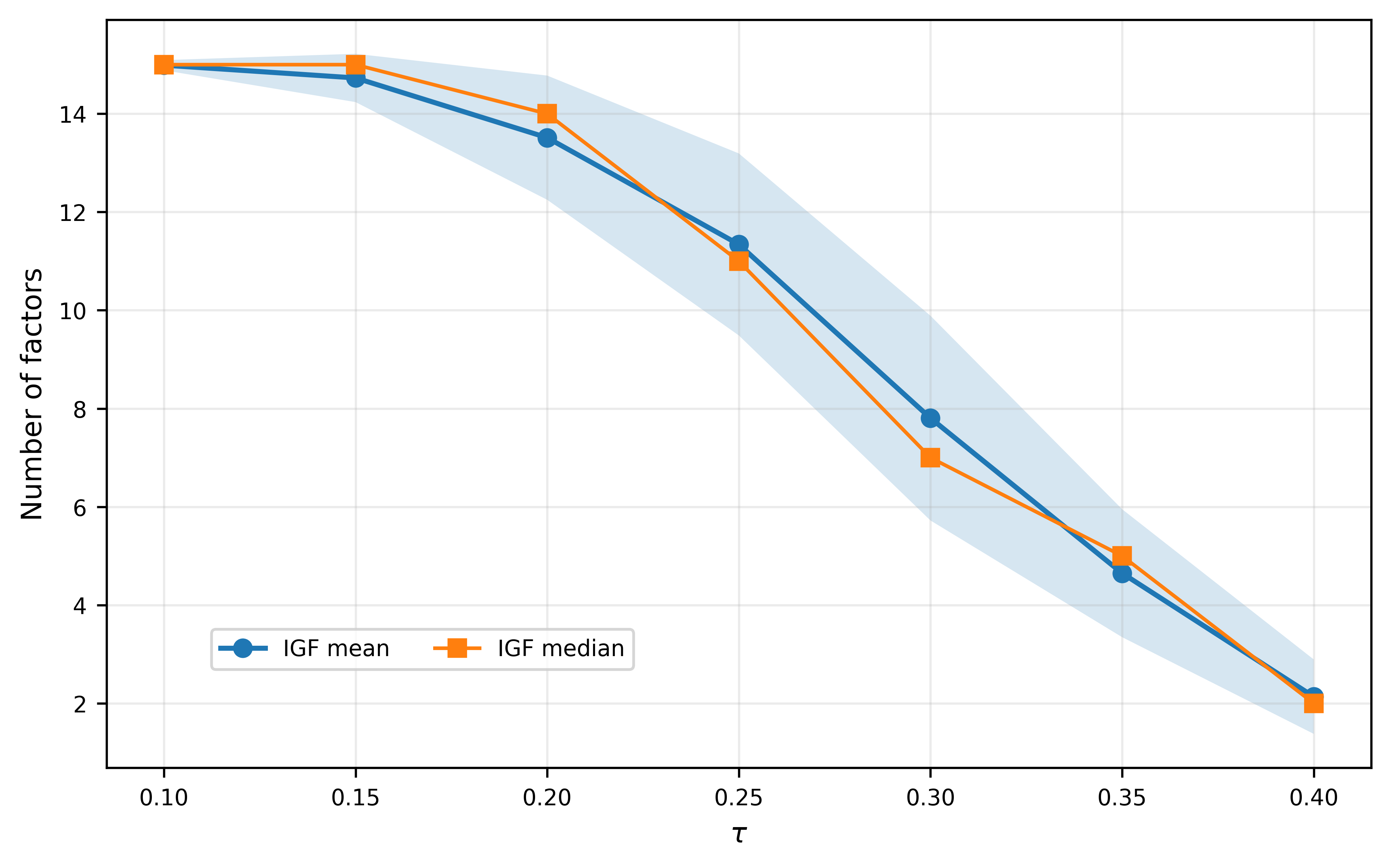}
    \caption{Sensitivity of the IGF factor count to the participation threshold $\tau$ in the empirical moving-window analysis.}
    \label{fig:empirical_windows_sensitivity}
\end{figure}

Figure~\ref{fig:empirical_windows_sensitivity} shows how the mean and median number of factors detected by the IGF algorithm change with the threshold $\tau$ across the $m=185$ moving windows. It is worth noting that, even for large values of $\tau$, the mean factor count remains above the one-factor detection level of the Onatski test. Thus, even when retaining only highly extensive factors, the IGF algorithm still detects, on average, more than one factor.

\section{Conclusion}

Our analysis shows that the IGF algorithm can recover the true number of factors in the BH factor model, whereas state-of-the-art eigenvalue-based criteria fail to recover it in the same regime. Moreover, by calibrating the local decision threshold $\tau$ in a synthetic factor model matched to the empirical dimensions, the same procedure can be applied to real data in a controlled way. In the empirical S\&P 500 analysis, the detected number of global factors exhibits nontrivial temporal variation, with a median count of 7 factors. This number is substantially larger than the typical one-factor detection obtained from the Onatski test.

The factors estimated by IGF are statistical factors and should not be interpreted directly as economic factors. Rather, they provide a first screening criterion for identifying candidate global directions in the return correlation matrix. A natural next step is to regress these statistical factors against established economic factors, such as those associated with the capital asset pricing model \cite{sharpe1964capital}, the Fama--French factors \cite{fama1993common}, and the broader factor zoo discussed in Ref.~\cite{harvey2016and}.

In this sense, the IGF algorithm acts as a discriminating criterion for selecting factors with global and extensive eigenvector structure. The PR filter helps exclude localized directions that may be statistically separated from the bulk but are not broadly distributed across assets. This is particularly relevant in the high-dimensional regime and near the BBP transition, where weak factors compete with fluctuations at the Marčenko--Pastur spectral edge and eigenvalue-only criteria may become ambiguous.

A further contribution of this work is the derivation of asymptotic PR benchmarks for the BH factor model. We show that weak-factor directions and typical idiosyncratic sample eigenvectors satisfy the delocalized limit $\mathrm{PR}(u)/p\to 1/3$. Equivalently, this corresponds to the inverse-participation benchmark $p\sum_i u_i^4\to 3$, previously obtained for delocalized graph-Laplacian eigenvectors \cite{clark2018moments}. By contrast, the leading coherent common eigenvector satisfies $\mathrm{PR}(u_1)/p\to 1$ in the strong common-loading regime. These limits provide an eigenvector-level criterion for distinguishing localized components from extensive global factors.

Several extensions remain open. First, the threshold $\tau=0.3$ should be understood as an operational calibration for the empirical dimensions studied here, rather than as a universal constant. Second, future work should examine the stability and economic interpretation of the detected factors across different markets, sampling frequencies, and time periods. Finally, an important theoretical direction is the development of operational methods for detecting subcritical factors hidden inside the Marčenko--Pastur bulk, possibly by exploiting information from the full eigenvalue spectrum through free-probability techniques.
%%%

\medskip
\noindent{\bf Competing Interests:} The authors have no competing interests to declare that are relevant to the content of this article.
\medskip

\noindent{\bf Data availability statement:} The data that support the findings of this study are available from the corresponding author upon request.
\medskip

\noindent
{\bf Declaration of AI-assisted work.}
The author used \texttt{GPT-5.5 Thinking} to assist with language editing, code review, and mathematical exposition. All AI-assisted material was reviewed, revised, and verified by the author, who remains fully responsible for the manuscript.
\medskip

\noindent
{\bf Author Contribution Statement.}
The author A.G.M. confirms being the sole contributor to this work and has approved it for publication.
The author was responsible for the conception and design of the study, data collection, analysis, interpretation of results, and the writing and revision of the manuscript.

%%%
\appendix
\renewcommand{\thesection}{\Alph{section}}

\section{Critical values of the Onatski test}
\label{appendix:a}
Table~\ref{a1} show the significance levels for the R~ statistic (for computational details see \cite{onatski2009testing}).
\begin{table}[htbp]
\centering
\begin{tabular}{|r|r|r|r|r|r|r|r|r|}
\hline
size & \multicolumn{8}{c}{$k_1-k_0$} \vline \\ \hline
\multicolumn{1}{|l|}{\%} & 1 & 2 & 3 & 4 & 5 & 6 & 7 & 8 \\ \hline
15 & 2.75 & 3.62 & 4.15 & 4.54 & 4.89 & 5.2 & 5.45 & 5.7 \\ \hline
10 & 3.33 & 4.31 & 4.91 & 5.4 & 5.77 & 6.13 & 6.42 & 6.66 \\ \hline
9 & 3.5 & 4.49 & 5.13 & 5.62 & 6.03 & 6.39 & 6.67 & 6.92 \\ \hline
8 & 3.69 & 4.72 & 5.37 & 5.91 & 6.31 & 6.68 & 6.95 & 7.25 \\ \hline
7 & 3.92 & 4.99 & 5.66 & 6.24 & 6.62 & 7 & 7.32 & 7.59 \\ \hline
6 & 4.2 & 5.31 & 6.03 & 6.57 & 7 & 7.41 & 7.74 & 8.04 \\ \hline
5 & 4.52 & 5.73 & 6.46 & 7.01 & 7.5 & 7.95 & 8.29 & 8.59 \\ \hline
4 & 5.02 & 6.26 & 6.97 & 7.63 & 8.16 & 8.61 & 9.06 & 9.36 \\ \hline
3 & 5.62 & 6.91 & 7.79 & 8.48 & 9.06 & 9.64 & 10.11 & 10.44 \\ \hline
2 & 6.55 & 8.15 & 9.06 & 9.93 & 10.47 & 11.27 & 11.75 & 12.13 \\ \hline
1 & 8.74 & 10.52 & 11.67 & 12.56 & 13.42 & 14.26 & 14.88 & 15.25 \\ \hline
\end{tabular}
\caption{The rows represent the level of significance, and the columns the size of the test.}
\label{a1}
\end{table}

\section{Coefficient of Variation of $D_{ii}$}
\label{appendix:b}
We have
\begin{equation}
    \ell_i^2
    =
    \sum_{\alpha=1}^{K}L_{i\alpha}^2,
\end{equation}
with
\begin{equation}
    L_{i\alpha}=b+\eta_{i\alpha},
    \qquad
    \eta_{i\alpha}\sim N(0,\sigma_b^2).
\end{equation}
Then $L_{i\alpha}\sim N(b,\sigma_b^2)$, the mean and variance of the iid factor loadings $l^2_s$ are given by

\begin{eqnarray}
    \mu(l^2_s) &=& \mathbb{E}
    (l_s^2)
    =
    \mathbb{E}
    \left[
    \sum_{\alpha=1}^{K}L_{i\alpha}^2
    \right]
    =
    \sum_{\alpha=1}^{K}
    \mathbb{E}
    \left[
    L_{i\alpha}^2
    \right]
    =
    K(b^2+\sigma_b^2),\\
    \sigma^2(l^2_s) &=&
    \operatorname{Var}
    \left(
    \sum_{\alpha=1}^{K}L_{i\alpha}^2
    \right)
    =
    K\operatorname{Var}(L_{i\alpha}^2)
    =
    K
    \left(
    2\sigma_b^4+4b^2\sigma_b^2
    \right).
\end{eqnarray}
where we have used 
\begin{equation}
    \mathbb{E}[L_{i\alpha}^2]=b^2+\sigma_b^2,
    \qquad
    \mathbb{E}[L_{i\alpha}^4]=b^4+6b^2\sigma_b^2+3\sigma_b^4.
\end{equation}
and,
\begin{align}
    \operatorname{Var}(L_{i\alpha}^2)
    &=
    \mathbb{E}[L_{i\alpha}^4]
    -
    \left(\mathbb{E}[L_{i\alpha}^2]\right)^2 \\
    &=
    b^4+6b^2\sigma_b^2+3\sigma_b^4
    -
    (b^2+\sigma_b^2)^2 \\
    &=
    2\sigma_b^4+4b^2\sigma_b^2.
\end{align}

Thus, the coefficient of variation of $D_{ii}$ is
\begin{equation}
    CV_D
    =
    \frac{
    \sigma_f^2
    \sqrt{
    K(2\sigma_b^4+4b^2\sigma_b^2)
    }
    }{
    \sigma_e^2+\sigma_f^2K(b^2+\sigma_b^2)
    }.
    \label{eq:cvD}
\end{equation}

\section{Participation ratio of BH model}
\label{appendix:c}

Consider the population covariance matrix of BH factor model
\begin{equation}
    \mathbf{\Sigma}_{BH}=\sigma_f^2 \mathbf{LL}^\top+\sigma_e^2 \mathbf{I} .
\end{equation}
An alternative matrix representation of the loadings is 
\begin{equation}
    \mathbf{L} = b\,\mathbf{1}_p \mathbf{1}_K^\top + \sigma_b \mathbf{Z},
    \qquad
    Z_{ik}\sim N(0,1),
    \label{e1_loadings}
\end{equation}
where $\mathbf{1}_p, \mathbf{1}_K$ represent a vectors of ones of dimension $p,K$, respectively, and $\mathbf{Z}$ is a random standard gaussian matrix of the same dimension as $\mathbf{L}$ (i.e. $p\times K$). Notice that the term $\sigma_e^2 \mathbf{I}$ only shifts the eigenvalues. Hence the
nontrivial population eigenvectors are determined by $\mathbf{LL}^\top$. 

The analysis can be carried out in terms of $\mathbf{LL}^\top$ because it
has the same eigenspace as $\mathbf{\Sigma}$. In other words, 
$\operatorname{rank}(\mathbf{LL}^\top)\leq K$, and only the first \(K\) directions are factor directions. The remaining \(p-K\) directions belong to the null
space of $\mathbf{L}^\top$. Then, the nonzero eigenvectors of  $\mathbf{LL}^\top$  can be obtained from the eigenvectors of  $\mathbf{L}^\top\mathbf{L}$. 
Substituting Eq.\ref {e1_loadings} into the normalized Gram matrix of the loadings. 
\begin{align}
    \frac{1}{p}\mathbf{L}^\top\mathbf{L}
    &=
    b^2\mathbf{1}_K\mathbf{1}_K^\top
    +
    \frac{b\sigma_b}{p}
    \left(
    \mathbf{1}_K\mathbf{1}_p^\top \mathbf{Z}
    +
    \mathbf{Z}^\top\mathbf{1}_p\mathbf{1}_K^\top
    \right)
    +
    \sigma_b^2\frac{1}{p}\mathbf{Z}^\top \mathbf{Z}.
\end{align}
This matrix converges to
\begin{equation}
    \frac{1}{p}\mathbf{L}^\top\mathbf{L}
    \to
    b^2\mathbf{1}_K\mathbf{1}_K^\top+\sigma_b^2\mathbf{I}_K\quad \text{as}\quad p\rightarrow \infty
\end{equation}

% because 
% \begin{equation}
%     \frac{1}{p}Z^\top Z\to I_K,
%     \qquad
%     \frac{1}{p}\mathbf{1}_p^\top Z\to 0 .
% \end{equation}

\subsection{Proof of $\mathrm{PR}(\mathbf{u}_1)/p\to 1$}
Let $\mathbf{G}$ denote the limiting normalized Gram matrix of the loadings,
\begin{equation}
    \mathbf{G}
    =
    b^2\mathbf{1}_K\mathbf{1}_K^\top
    +
    \sigma_b^2\mathbf{I}_K,
\end{equation}
The leading eigenpair of $\mathbf{G}$ is
\begin{equation}
    \mathbf{a}_1
    =
    \frac{\mathbf{1}_K}{\sqrt{K}},
    \qquad
    \gamma_1
    =
    Kb^2+\sigma_b^2,
\end{equation}
because
\begin{equation}
    \mathbf{G}\mathbf{a}_1
    =
    \gamma_1 \mathbf{a}_1.
\end{equation}

Then, the leading eigenvalue of $\mathbf{L}^\top\mathbf{L}$
\begin{equation}
    d_1
    \approx
    p\left(Kb^2+\sigma_b^2\right).
\end{equation}

On the other hand, the eigenvectors of $\mathbf{L}\mathbf{L}^\top$ associated with nonzero
eigenvalues are obtained from the eigenvectors of
$\mathbf{L}^\top\mathbf{L}$. In particular, if
\begin{equation}
    \mathbf{L}^\top\mathbf{L}\mathbf{a}_1=d_1\mathbf{a}_1,
    \qquad
    \|\mathbf{a}_1\|=1,
\end{equation}
then the corresponding normalized eigenvector of
$\mathbf{L}\mathbf{L}^\top$ is
\begin{equation}
    \mathbf{u}_1
    =
    \frac{\mathbf{L}\mathbf{a}_1}{\sqrt{d_1}}.
\end{equation}
We obtain
\begin{align}
    \mathbf{L}\mathbf{a}_1
    &=
    b\,\mathbf{1}_p\mathbf{1}_K^\top
    \frac{\mathbf{1}_K}{\sqrt{K}}
    +
    \sigma_b\mathbf{Z}
    \frac{\mathbf{1}_K}{\sqrt{K}} \\
    &=
    b\sqrt{K}\,\mathbf{1}_p
    +
    \sigma_b\mathbf{g}_1,
\end{align}
where
\begin{equation}
    \mathbf{g}_1
    =
    \mathbf{Z}\frac{\mathbf{1}_K}{\sqrt{K}}\quad
    \mathbf{g}_1\sim N\left(\mathbf{0},\mathbf{I}_p\right).
\end{equation}

Therefore, for fixed \(K\) and large \(p\),
\begin{equation}
    \mathbf{u}_1
    \approx
    \frac{
    b\sqrt K\,\mathbf{1}_p+\sigma_b \mathbf{g}_1
    }
    {
    \sqrt{p(b^2K+\sigma_b^2)}
    }.
\end{equation}

In the strong common-loading regime $b^2K\gg \sigma_b^2$, 
\begin{equation}
    \mathbf{u}_1
    \approx
    \frac{\mathbf{1}_p}{\sqrt p}.
\end{equation}
Consequently,
\begin{equation}
    \frac{\operatorname{PR}(\mathbf{u}_1)}{p}
    \to
    1.
\end{equation}

\subsection{Proof of $\mathrm{PR}(\mathbf{u}_\alpha)\rightarrow 1/3$, for $\alpha=2,\dots, K$}

Now consider the remaining factor-space directions. Let
$\mathbf{a}_\alpha\in\mathbb{R}^K$, $\alpha=2,\dots,K$, be unit vectors
orthogonal to $\mathbf{1}_K$, that is,
\begin{equation}
    \|\mathbf{a}_\alpha\|=1,
    \qquad
    \mathbf{1}_K^\top \mathbf{a}_\alpha=0,
    \qquad
    \alpha=2,\dots,K.
\end{equation}
Then
\begin{equation}
    \mathbf{G}\mathbf{a}_\alpha
    =
    \sigma_b^2\mathbf{a}_\alpha,
    \qquad
    \alpha=2,\dots,K.
\end{equation}
Therefore, if $d_\alpha$ denotes the corresponding eigenvalue of
$\mathbf{L}^\top\mathbf{L}$, then
\begin{equation}
    d_\alpha
    \approx
    p\sigma_b^2,
    \qquad
    \alpha=2,\dots,K.
\end{equation}

Using the singular-vector relation,
\begin{equation}
    \mathbf{u}_\alpha
    =
    \frac{\mathbf{L}\mathbf{a}_\alpha}{\sqrt{d_\alpha}},
    \qquad
    \alpha=2,\dots,K,
\end{equation}
we obtain
\begin{equation}
    \mathbf{L}\mathbf{a}_\alpha
    =
    b\,\mathbf{1}_p\mathbf{1}_K^\top\mathbf{a}_\alpha
    +
    \sigma_b\mathbf{Z}\mathbf{a}_\alpha 
    =
    \sigma_b\mathbf{Z}\mathbf{a}_\alpha
    = 
    \sigma_b \mathbf{g}_\alpha,
\end{equation}
because $\mathbf{1}_K^\top\mathbf{a}_\alpha=0$, and has been defined
\begin{equation}
    \mathbf{g}_\alpha
    =
    \mathbf{Z}\mathbf{a}_\alpha,
    \qquad
    \mathbf{g}_\alpha\sim N\left(\mathbf{0},\mathbf{I}_p\right).
\end{equation}
We get
\begin{equation}
    \mathbf{u}_\alpha
    \approx
    \frac{\mathbf{g}_\alpha}{\sqrt{p}},
    \qquad
    \alpha=2,\dots,K.
\end{equation}

Therefore,
\begin{equation}
    \frac{\operatorname{PR}(\mathbf{u}_\alpha)}{p}
    =
    \frac{1}
    {p\sum_{i=1}^{p}u_{\alpha i}^4}
    \approx
    \frac{1}
    {
    \frac{1}{p}
    \sum_{i=1}^{p}g_{\alpha i}^4.}
\end{equation}
By the law of large numbers,
\begin{equation}
    p^{-1}\sum_{i=1}^{p}g_{\alpha i}^4
    \to
    3
\end{equation}
Hence,
\begin{equation}
    \frac{\operatorname{PR}(\mathbf{u}_\alpha)}{p}
    \to
    \frac{1}{3},
    \qquad
    \alpha=2,\dots,K.
\end{equation}

\subsection{Proof of $\mathrm{PR}(\mathbf{u}_{\alpha})\rightarrow 1/3$, for $\alpha>K$}

Finally, the remaining directions satisfy
\begin{equation}
    \mathbf{L}^\top\mathbf{u}_\alpha=0,
    \qquad
    \alpha>K.
\end{equation}
Hence
\begin{equation}
    \mathbf{L}\mathbf{L}^\top\mathbf{u}_\alpha=0.
\end{equation}
For the population covariance matrix
\begin{equation}
    \mathbf{\Sigma}_{BH}
    =
    \sigma_f^2\mathbf{L}\mathbf{L}^\top
    +
    \sigma_e^2\mathbf{I}_p,
\end{equation}
we obtain
\begin{equation}
    \mathbf{\Sigma}_{BH}\mathbf{u}_\alpha
    =
    \sigma_e^2\mathbf{u}_\alpha,
    \qquad
    \alpha>K.
\end{equation}
Therefore, the idiosyncratic eigenspace of $\mathbf{\Sigma}_{BH}$ is
$\operatorname{Null}(\mathbf{L}^\top)$, with eigenvalue $\sigma_e^2$.

At the population level, this eigenvalue has multiplicity $p-K$. Hence the
individual eigenvectors inside $\operatorname{Null}(\mathbf{L}^\top)$ are
not uniquely defined. Any orthonormal basis of this subspace is admissible,
so the population model does not assign a unique value of
$\operatorname{PR}(\mathbf{u}_\alpha)/p$ to each idiosyncratic eigenvector.

In finite samples, this degeneracy is broken by random fluctuations. Thus we write
\begin{equation}
    \mathbf{u}_\alpha
    \approx
    \frac{\mathbf{g}_\alpha}{\|\mathbf{g}_\alpha\|},
    \qquad
    \mathbf{g}_\alpha\sim N\left(\mathbf{0},\mathbf{I}_p\right),
    \qquad
    \alpha>K.
\end{equation}
Hence,
\begin{equation}
    \frac{\operatorname{PR}(\mathbf{u}_\alpha)}{p}
    \approx
    \frac{1}{3},
    \qquad
    \alpha>K.
\end{equation}

\subsection{$\mathrm{PR}$ limits for the correlation matrix}
\label{appendix:c4}
When $CV_D\ll 1$, the diagonal entries are close to their typical scale,
so that
\begin{equation}
    \mathbf{D}
    \approx
    d_0\mathbf{I}_p.
\end{equation}
Hence
\begin{equation}
    \mathbf{C}_{BH}
    \approx
    \frac{1}{d_0}\mathbf{\Sigma}_{BH}
    =
    \sigma_f^2\widetilde{\mathbf{L}}\widetilde{\mathbf{L}}^\top
    +
    \sigma_0^2\mathbf{I}_p,
\end{equation}
where
\begin{equation}
    \widetilde{\mathbf{L}}
    =
    \frac{\mathbf{L}}{\sqrt{d_0}},
    \qquad
    \sigma_0^2
    =
    \frac{\sigma_e^2}{d_0}.
\end{equation}

The normalized loading Gram matrix in correlation scale is
\begin{equation}
    \frac{1}{p}
    \widetilde{\mathbf{L}}^\top
    \widetilde{\mathbf{L}}
    =
    \frac{1}{d_0}
    \frac{1}{p}
    \mathbf{L}^\top\mathbf{L}.
\end{equation}
Therefore, the eigenvalues are rescaled by $d_0^{-1}$, but the
factor-space eigenvectors are unchanged.

If $\mathbf{a}_\alpha$ is a factor-space direction, then the corresponding
correlation-scale eigenvector is proportional to
\begin{equation}
    \widetilde{\mathbf{L}}\mathbf{a}_\alpha
    =
    \frac{1}{\sqrt{d_0}}
    \mathbf{L}\mathbf{a}_\alpha.
\end{equation}
After normalization, the scalar factor cancels:
\begin{equation}
    \mathbf{u}_\alpha^{(C)}
    =
    \frac{
    \widetilde{\mathbf{L}}\mathbf{a}_\alpha
    }{
    \|\widetilde{\mathbf{L}}\mathbf{a}_\alpha\|
    }
    =
    \frac{
    d_0^{-1/2}\mathbf{L}\mathbf{a}_\alpha
    }{
    \|d_0^{-1/2}\mathbf{L}\mathbf{a}_\alpha\|
    }
    =
    \frac{
    \mathbf{L}\mathbf{a}_\alpha
    }{
    \|\mathbf{L}\mathbf{a}_\alpha\|
    }
    =
    \mathbf{u}_\alpha^{(\Sigma)}.
\end{equation}
Thus, in the small-$CV_D$ regime, the correlation normalization changes the
eigenvalue scale but approximately preserves the eigenvector geometry.
Consequently, the same participation-ratio benchmarks apply. For the
leading common factor, in the strong common-loading regime
$Kb^2\gg\sigma_b^2$, and when the cross-sectional dispersion of the
diagonal variances is negligible
$CV_D\ll 1$
\begin{equation}
    \frac{\operatorname{PR}(\mathbf{u}_1^{(C)})}{p}
    \to
    1.
\end{equation}
For the weak factor directions,
\begin{equation}
    \frac{\operatorname{PR}(\mathbf{u}_\alpha^{(C)})}{p}
    \to
    \frac{1}{3},
    \qquad
    \alpha=2,\dots,K.
\end{equation}
For typical idiosyncratic sample eigenvectors,
\begin{equation}
    \frac{\operatorname{PR}(\mathbf{u}_\alpha^{(C)})}{p}
    \approx
    \frac{1}{3},
    \qquad
    \alpha>K.
\end{equation}

\bibliography{references}

@book{mehta2004random,
  title={Random matrices},
  author={Mehta, Madan Lal},
  volume={142},
  year={2004},
  publisher={Elsevier}
}

@article{laloux1999noise,
  title={Noise dressing of financial correlation matrices},
  author={Laloux, Laurent and Cizeau, Pierre and Bouchaud, Jean-Philippe and Potters, Marc},
  journal={Physical review letters},
  volume={83},
  number={7},
  pages={1467},
  year={1999},
  publisher={APS}
}

@article{plerou1999universal,
  title={Universal and nonuniversal properties of cross correlations in financial time series},
  author={Plerou, Vasiliki and Gopikrishnan, Parameswaran and Rosenow, Bernd and Amaral, Lu{\'\i}s A Nunes and Stanley, H Eugene},
  journal={Physical review letters},
  volume={83},
  number={7},
  pages={1471},
  year={1999},
  publisher={APS}
}

@article{benaych2011eigenvalues,
  title={The eigenvalues and eigenvectors of finite, low rank perturbations of large random matrices},
  author={Benaych-Georges, Florent and Nadakuditi, Raj Rao},
  journal={Advances in Mathematics},
  volume={227},
  number={1},
  pages={494--521},
  year={2011},
  publisher={Elsevier}
}

@article{benaych2012singular,
  title={The singular values and vectors of low rank perturbations of large rectangular random matrices},
  author={Benaych-Georges, Florent and Nadakuditi, Raj Rao},
  journal={Journal of Multivariate Analysis},
  volume={111},
  pages={120--135},
  year={2012},
  publisher={Elsevier}
}

@article{gavish2014optimal,
  title={The optimal hard threshold for singular values is $4/\sqrt{3}$},
  author={Gavish, Matan and Donoho, David L},
  journal={IEEE Transactions on Information Theory},
  volume={60},
  number={8},
  pages={5040--5053},
  year={2014},
  publisher={IEEE}
}

@article{marvcenko1967distribution,
  title={Distribution of eigenvalues for some sets of random matrices},
  author={Mar{\v{c}}enko, Vladimir A and Pastur, Leonid Andreevich},
  journal={Mathematics of the USSR-Sbornik},
  volume={1},
  number={4},
  pages={457},
  year={1967},
  publisher={IOP Publishing}
}

@article{onatski2008tracy,
  title={The Tracy--Widom limit for the largest eigenvalues of singular complex Wishart matrices},
  author={Onatski, Alexei},
  journal={The Annals of Applied Probability},
  volume={18},
  number={2},
  pages={470--490},
  year={2008},
  publisher={Institute of Mathematical Statistics}
}

@article{tracy1994level,
  title={Level-spacing distributions and the Airy kernel},
  author={Tracy, Craig A and Widom, Harold},
  journal={Communications in Mathematical Physics},
  volume={159},
  number={1},
  pages={151--174},
  year={1994},
  publisher={Springer}
}

@article{onatski2009testing,
  title={Testing hypotheses about the number of factors in large factor models},
  author={Onatski, Alexei},
  journal={Econometrica},
  volume={77},
  number={5},
  pages={1447--1479},
  year={2009},
  publisher={Wiley Online Library}
}

@article{forni2000generalized,
  title={The generalized dynamic-factor model: Identification and estimation},
  author={Forni, Mario and Hallin, Marc and Lippi, Marco and Reichlin, Lucrezia},
  journal={Review of Economics and statistics},
  volume={82},
  number={4},
  pages={540--554},
  year={2000},
  publisher={MIT Press 238 Main St., Suite 500, Cambridge, MA 02142-1046, USA journals}
}

@article{plerou2002random,
  title={Random matrix approach to cross correlations in financial data},
  author={Plerou, Vasiliki and Gopikrishnan, Parameswaran and Rosenow, Bernd and Amaral, Luis A Nunes and Guhr, Thomas and Stanley, H Eugene},
  journal={Physical Review E},
  volume={65},
  number={6},
  pages={066126},
  year={2002},
  publisher={APS}
}

@inproceedings{Iain2006,
  title={High dimensional statistical inference and random matrices},
  author={Johnstone, Iain M.},
  booktitle={Proceedings of the International Congress of Mathematicians},
  pages={307--333},
  year={2006},
  address={Madrid, Spain}
}

@article{harding2008explaining,
  title={Explaining the single factor bias of arbitrage pricing models in finite samples},
  author={Harding, Matthew C},
  journal={Economics Letters},
  volume={99},
  number={1},
  pages={85--88},
  year={2008},
  publisher={Elsevier}
}

@article{kritchman2009non,
  title={Non-parametric detection of the number of signals: Hypothesis testing and random matrix theory},
  author={Kritchman, Shira and Nadler, Boaz},
  journal={IEEE Transactions on Signal Processing},
  volume={57},
  number={10},
  pages={3930--3941},
  year={2009},
  publisher={IEEE}
}

@article{molero2023market,
  title={Market beta is not dead: An approach from random matrix theory},
  author={Molero-Gonz{\'a}lez, L and Trinidad-Segovia, JE and S{\'a}nchez-Granero, MA and Garc{\'\i}a-Medina, A},
  journal={Finance Research Letters},
  volume={55},
  pages={103816},
  year={2023},
  publisher={Elsevier}
}

@article{sharpe1964capital,
  title={Capital asset prices: A theory of market equilibrium under conditions of risk},
  author={Sharpe, William F},
  journal={The journal of finance},
  volume={19},
  number={3},
  pages={425--442},
  year={1964},
  publisher={Wiley Online Library}
}

@article{fama1993common,
  title={Common risk factors in the returns on stocks and bonds},
  author={Fama, Eugene F and French, Kenneth R},
  journal={Journal of financial economics},
  volume={33},
  number={1},
  pages={3--56},
  year={1993},
  publisher={Elsevier}
}

@article{harvey2016and,
  title={… and the cross-section of expected returns},
  author={Harvey, Campbell R and Liu, Yan and Zhu, Heqing},
  journal={The Review of financial studies},
  volume={29},
  number={1},
  pages={5--68},
  year={2016},
  publisher={Oxford University Press}
}

@article{cattell1966scree,
  title={The scree test for the number of factors},
  author={Cattell, Raymond B},
  journal={Multivariate behavioral research},
  volume={1},
  number={2},
  pages={245--276},
  year={1966},
  publisher={Taylor \& Francis}
}

@article{wold1978cross,
  title={Cross-validatory estimation of the number of components in factor and principal components models},
  author={Wold, Svante},
  journal={Technometrics},
  volume={20},
  number={4},
  pages={397--405},
  year={1978},
  publisher={Taylor \& Francis}
}

@article{zientek2007applying,
  title={Applying the bootstrap to the multivariate case: Bootstrap component/factor analysis},
  author={Zientek, Linda Reichwein and Thompson, Bruce},
  journal={Behavior research methods},
  volume={39},
  number={2},
  pages={318--325},
  year={2007},
  publisher={Springer}
}

@article{kaiser1960application,
  title={The application of electronic computers to factor analysis},
  author={Kaiser, Henry F},
  journal={Educational and psychological measurement},
  volume={20},
  number={1},
  pages={141--151},
  year={1960},
  publisher={Sage Publications Sage CA: Thousand Oaks, CA}
}

@article{guttman1954some,
  title={Some necessary conditions for common-factor analysis},
  author={Guttman, Louis},
  journal={Psychometrika},
  volume={19},
  number={2},
  pages={149--161},
  year={1954},
  publisher={Springer-Verlag}
}

@article{onatski2010determining,
  title={Determining the number of factors from empirical distribution of eigenvalues},
  author={Onatski, Alexei},
  journal={The Review of Economics and Statistics},
  volume={92},
  number={4},
  pages={1004--1016},
  year={2010},
  publisher={The MIT Press}
}

@article{bai2002determining,
  title={Determining the number of factors in approximate factor models},
  author={Bai, Jushan and Ng, Serena},
  journal={Econometrica},
  volume={70},
  number={1},
  pages={191--221},
  year={2002},
  publisher={Wiley Online Library}
}

@article{onatski2012asymptotics,
  title={Asymptotics of the principal components estimator of large factor models with weakly influential factors},
  author={Onatski, Alexei},
  journal={Journal of Econometrics},
  volume={168},
  number={2},
  pages={244--258},
  year={2012},
  publisher={Elsevier}
}

@article{johnstone2001distribution,
  title={On the distribution of the largest eigenvalue in principal components analysis},
  author={Johnstone, Iain M},
  journal={The Annals of statistics},
  volume={29},
  number={2},
  pages={295--327},
  year={2001},
  publisher={Institute of Mathematical Statistics}
}

@article{baik2005phase,
  title   = {Phase transition of the largest eigenvalue for nonnull complex sample covariance matrices},
  author  = {Baik, Jinho and Ben Arous, G{\'e}rard and P{\'e}ch{\'e}, Sandrine},
  journal = {Annals of Probability},
  volume  = {33},
  number  = {5},
  pages   = {1643--1697},
  year    = {2005},
  publisher = {Institute of Mathematical Statistics}
}

@article{paul2007asymptotics,
  title={Asymptotics of sample eigenstructure for a large dimensional spiked covariance model},
  author={Paul, Debashis},
  journal={Statistica Sinica},
  pages={1617--1642},
  year={2007},
  publisher={JSTOR}
}

@article{baik2006eigenvalues,
  title={Eigenvalues of large sample covariance matrices of spiked population models},
  author={Baik, Jinho and Silverstein, Jack W},
  journal={Journal of multivariate analysis},
  volume={97},
  number={6},
  pages={1382--1408},
  year={2006},
  publisher={Elsevier}
}

@article{johnstone2020testing,
  title   = {Testing in High-Dimensional Spiked Models},
  author  = {Johnstone, Iain M. and Onatski, Alexei},
  journal = {The Annals of Statistics},
  year    = {2020},
  volume  = {48},
  number  = {3},
  pages   = {1231--1254},
  publisher = {Institute of Mathematical Statistics}
}

@article{brown1989number,
  title={The number of factors in security returns},
  author={Brown, Stephen J},
  journal={the Journal of Finance},
  volume={44},
  number={5},
  pages={1247--1262},
  year={1989},
  publisher={Wiley Online Library}
}

@phdthesis{harding2007essays,
  title  = {Essays in Econometrics and Random Matrix Theory},
  author = {Harding, Matthew C.},
  school = {Massachusetts Institute of Technology},
  address = {Cambridge, MA},
  year   = {2007},
  type   = {Ph.D. dissertation},
  note   = {Department of Economics},
  url    = {https://hdl.handle.net/1721.1/39670}
}

@misc{yeo2016random,
  title         = {Random Matrix Approach to Estimation of High-Dimensional Factor Models},
  author        = {Yeo, Joongyeub and Papanicolaou, George},
  year          = {2016},
  eprint        = {1611.05571},
  archivePrefix = {arXiv},
  primaryClass  = {math.ST}
}

@article{clark2018moments,
  title={Moments of the inverse participation ratio for the Laplacian on finite regular graphs},
  author={Clark, Timothy BP and Del Maestro, Adrian},
  journal={Journal of Physics A: Mathematical and Theoretical},
  volume={51},
  number={49},
  pages={495003},
  year={2018},
  publisher={IOP Publishing}
}

@article{johnstone2008multivariate,
  title={Multivariate analysis and Jacobi ensembles: Largest eigenvalue, Tracy--Widom limits and rates of convergence},
  author={Johnstone, Iain M},
  journal={Annals of statistics},
  volume={36},
  number={6},
  pages={2638},
  year={2008}
}

@Article{harris2020array,
 title         = {Array programming with {NumPy}},
 author        = {Charles R. Harris and K. Jarrod Millman and St{\'{e}}fan J.
                 van der Walt and Ralf Gommers and Pauli Virtanen and David
                 Cournapeau and Eric Wieser and Julian Taylor and Sebastian
                 Berg and Nathaniel J. Smith and Robert Kern and Matti Picus
                 and Stephan Hoyer and Marten H. van Kerkwijk and Matthew
                 Brett and Allan Haldane and Jaime Fern{\'{a}}ndez del
                 R{\'{i}}o and Mark Wiebe and Pearu Peterson and Pierre
                 G{\'{e}}rard-Marchant and Kevin Sheppard and Tyler Reddy and
                 Warren Weckesser and Hameer Abbasi and Christoph Gohlke and
                 Travis E. Oliphant},
 year          = {2020},
 month         = sep,
 journal       = {Nature},
 volume        = {585},
 number        = {7825},
 pages         = {357--362},
 publisher     = {Springer Science and Business Media {LLC}},
 url           = {https://doi.org/10.1038/s41586-020-2649-2}
}
\end{document}